\documentclass[reqno]{amsart}

\usepackage[breakable,skins]{tcolorbox}
\usepackage{hyperref}
\hypersetup{pdfauthor=author}
\usepackage{graphicx,cuted,xcolor}  \usepackage{bbding} 
\usepackage{amssymb,latexsym,amsfonts,amsmath,amsthm}
\usepackage{diagrams}
\usepackage{arydshln}
\usepackage{verbatim}
\usepackage{temporal}
\usepackage{comment}
\usepackage{booktabs} 
\usepackage{amssymb,latexsym,amsfonts,amsmath,amsthm}
\usepackage{url}
\usepackage{xurl}
\usepackage{color}
\usepackage{subfigure}
\usepackage{flushend}
\usepackage{enumerate}
\usepackage[normalem]{ulem}

\usepackage{array}
\usepackage{makecell}
\usepackage[justification=justified]{caption}
\allowdisplaybreaks

\def\rmdef{\stackrel{\mbox{\rm {\tiny def}}}{=}} 
\newcommand{\set}[1]{\left\{ #1 \right\}}

\def\boxspan{\mathit{span}}

\newcommand{\R}{{\mathbb{R}}}

\newcommand{\N}{{\mathbb{N}}}

\newtheorem{theorem}{Theorem}[section]

\newtheorem{problem}{Problem}

\newtheorem{definition}{Definition}
\newtheorem{example}[theorem]{Example}

\newtheorem{remark}[theorem]{Remark}

\newcommand{\seq}[1]{\langle #1 \rangle}

\usepackage[linesnumbered,ruled,algo2e]{algorithm2e}
\SetKwInput{KwInput}{Input}                
\SetKwInput{KwOutput}{Output}              

\usepackage{amssymb}
\usepackage{enumitem}
\setlist[enumerate,1]{label=\arabic*,leftmargin=2.5mm}
\setlist[enumerate,2]{label=\alph*),leftmargin=4.5mm}

\newcommand{\xhat}{\hat{x}}

\newtcolorbox{fancyquotes}{%
    enhanced jigsaw, 
    breakable,      
    frame hidden,   
    left=0.5cm,       
    right=0.5cm,      
    overlay={%
        \node [scale=6,
            text=black!80,
            inner sep=0.5pt,] at ([xshift=0cm,yshift=-0.7cm]frame.north west){``}; 
            },
                parbox=false,
}

\makeatletter
\long\def\@maketablecaption#1#2{\@tablecaptionsize
	\global \@minipagefalse
\hbox to \hsize{\parbox[t]{\hsize}{\centering #1 \\ #2}}}
\makeatother

\begin{document}

\begin{abstract}
Correct-by-construction synthesis is a cornerstone of the confluence of formal methods and control theory towards designing safety-critical systems. 
Instead of following the time-tested, albeit laborious  (re)design-verify-validate loop, correct-by-construction methodology advocates the use of continual refinements of formal requirements---connected by chains of formal proofs---to build a system that assures the correctness by design.  
A remarkable progress has been made in scaling the scope of applicability of correct-by-construction synthesis---with a focus on cyber-physical systems that tie discrete-event control with continuous environment---to enlarge control systems by combining symbolic approaches with principled state-space reduction techniques.
Unfortunately, in the security-critical control systems, the security properties are verified \emph{ex post facto} the design process in a way that undermines the correct-by-construction paradigm. We posit that, to truly realize the dream of correct-by-construction synthesis for security-critical systems, security considerations must take center-stage with the safety considerations. 
Moreover, catalyzed by the recent progress on the opacity sub-classes of security properties and the notion of hyperproperties capable of combining security with safety properties, we believe that the time is ripe for the research community to holistically target the challenge of \emph{secure-by-construction} synthesis.
This paper details our vision by highlighting the recent progress and open challenges that may serve as bricks for providing a solid foundation for secure-by-construction synthesis of cyber-physical systems. 
\end{abstract}

\title[Secure-by-Construction Synthesis of Cyber-Physical Systems]{Secure-by-Construction Synthesis of Cyber-Physical Systems}

\author{Siyuan Liu$^{1,2}$}
\author{Ashutosh Trivedi$^3$}
\author{Xiang Yin$^4$}
\author{Majid Zamani$^{3,2}$}
\address{$^1$Department of Electrical and Computer Engineering, Technical University of Munich, 80333 Munich, Germany}
\email{sy.liu@tum.de} 
\address{$^2$Department of Computer Science, LMU Munich, 80538 Munich, Germany}
\email{}
\address{$^3$Department of Computer Science, University of Colorado Boulder, CO 80309, USA}
\email{ashutosh.trivedi@colorado.edu,majid.zamani@colorado.edu}
\address{$^4$Department of Automation, Shanghai Jiao Tong
	University, Shanghai 200240, China}
\email{yinxiang@sjtu.edu.cn}
\maketitle




The revolution in miniaturized communication devices in the beginning of this millennium contributed towards a revolution in the internet-of-things (IoT) and the networked systems woven around them: the cyber-physical systems (CPS).  
CPS are marked by a close-knit interaction of discrete computation and continuous control over a network and are playing critical roles in virtually every aspect of our modern experience ranging from consumer electronics to implantable medical devices, from smart cars to smart hospitals, and from controlling our power systems to safeguarding our nuclear rectors.
These systems are clearly safety-critical as a bug in their design could be life threatening, but given their societal implications, they are also security-critical where a bug in their design may have the potential to jeopardize the privacy, trust, and economic interests of society built around them. 

\begin{fancyquotes}
\emph{We believe that the security considerations should be elevated as primary design drivers along with safety ones to tackle the design challenge of modern CPS and call for a need to expand the correct-by-construction paradigm of designing safety-critical systems to encompass security: we call this paradigm \emph{secure-by-construction}.}
\end{fancyquotes}
This paper synthesizes ideas from three research communities: discrete event systems (DES), control systems (CS), and formal methods (FM) to pose and study central problems supporting secure-by-construction synthesis.

\section{Introduction}
 \begin{figure*}[ht!]
\begin{minipage}{0.635\textwidth}
\scriptsize
 \centering
 \hspace{-2cm}
\caption*{\tiny Services provided by average consultation, examination, and wait times}
 \begin{tabular}{ccccccc}
 \toprule
  \textbf{Service} & \makecell{ \textbf{Avg.} \\ \textbf{Total} \\ \textbf{Time} \\ \textbf{(min)}}  & \makecell{ \textbf{Avg.} \\ \textbf{Total}  \\ \textbf{Wait} \\ \textbf{(min)}}  & \makecell{ \textbf{Avg. Time} \\ \textbf{with} \\ \textbf{Nurse} \\ \textbf{(min)}}   & \makecell{ \textbf{Avg. Time} \\ \textbf{with} \\    \textbf{Physician} \\ \textbf{(min)}}  & \makecell{ \textbf{No.} \\ \textbf{Cases} \\  \textbf{(n)}} &  \makecell{ \textbf{Incom-} \\ \textbf{plete} \\ \textbf{Cases} \\ \textbf{(n)}} \\
   \midrule
\makecell{Minor \\ assessment \\  (std.)}  & 50 (30) & 33 (22) & 1 (3) &  16 (13) & 67 & 11 \\ 
 \makecell{Intermediate \\ assessment \\  (std.)}  & 55 (24) & 37 (21) & 2 (3) &  16 (12) & 400 & 29\\
 \makecell{General \\ assessment \\  (std.)}  & 77 (27) & 31 (17) & 10 (5) &  36 (19) & 43 & 1 \\
 \makecell{Psycho-\\therapy \\ (std.)}  & 71 (22) & 35 (16) & 2 (3) &  34 (14) & 11 & 0 \\
  \makecell{Annual exam \\ (after $16^{th}$ \\ birthday) (std.)}  & 51 (30) & 26 (12) & 7 (4) &  18 (8) & 5 & 2 \\
     \midrule
  \makecell{Other service}  &   &   &  &   &13 & N/A \\
   \makecell{No service \\ code given}  &   &   &  &   & 74 & N/A \\
  \bottomrule
 \end{tabular}
    \end{minipage}
    \hfill
    \begin{minipage}{0.345\textwidth}
\hspace{-.8cm}
\includegraphics[width=1.2\textwidth]{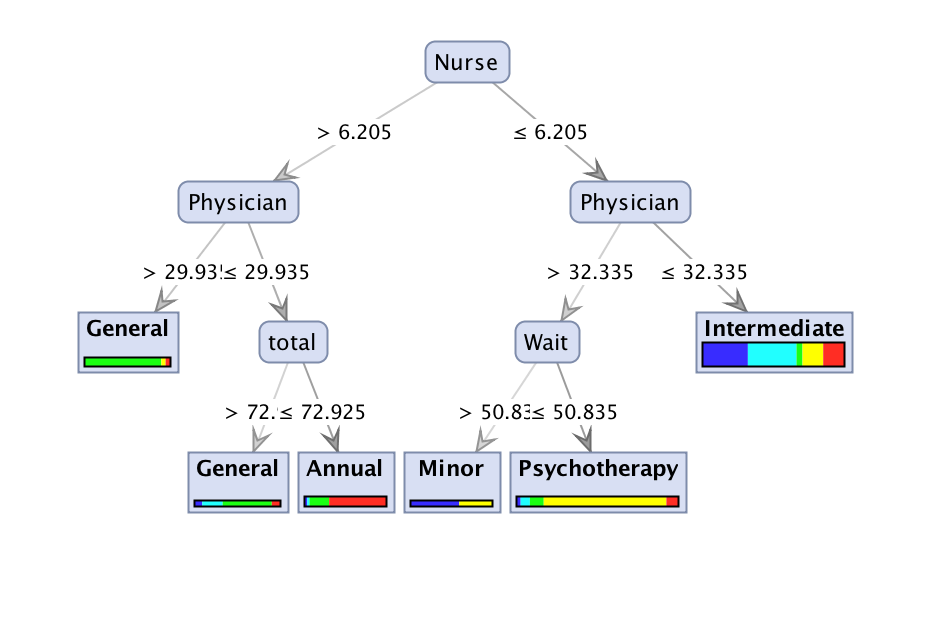}
\end{minipage}
   \caption{\small 
    Consider the dataset studied by Bestvater et al.~\cite{BDTN88},
    where the authors focused on the impact of waiting times on patient's perception
    of service satisfaction.
   This survey 
   collected the average time 
   patients spend with the nurse and the physician for various services ranging from major and minor assessments
    to psychotherapy.
    We emphasize that the dataset was carefully curated to minimize leaking
    any \emph{differentially private information} about the patients taking part in
    the survey. On the other hand,
    using a simple decision-tree classifier over this data, we found out that
    the timing data collected is leaking private information about patients in
    timing side-channels.
    For instance, if a patient spends less than $6$ minutes with the nurse
    and spends close to $32$ minutes with the physician with a low waiting time,
    the patient is visiting the hospital for a \emph{psychotherapy} session!}
    \label{fig:smart-hospital}
      \vspace{-0.3cm}
    \end{figure*}

Security considerations in the traditional computer science literature are often classified along the CIA mnemonic: \emph{confidentiality}, \emph{integrity}, and \emph{availability}. 
The confidentiality properties concern the protection of sensitive information leakage either directly or, more importantly, via side-channels (seemingly harmless observations of the system by malintent eavesdroppers). 
The umbrella-term integrity targets the establishment of the trust in the authenticity of the source of the information. Finally, availability properties concern with the protection of the system operations from cyberattacks aimed at disrupting or interrupting the core functionality of the system. 
While ensuring integrity deals with similar issues as for classical computer systems and can benefit from current best practices on encryption, the confidentiality and availability concerns in CPS get amplified due to a plethora of attack surfaces available in the form of physical system observations and constraints ranging from the usual time and memory to temperature, acoustics, pressure, and electro-magnetic radiation.

On the positive side, since principled approaches to CPS modeling and analysis already embrace the integration of the encoding of physical variables and discrete control, the confidentiality and availability properties can be explicated during the design time to ensure a system that is not only functional, but also guarantees freedom from known vulnerabilities. 
This is primary tenet of our stance on CPS-security: the design of security-critical CPS must tackle both functionality and security challenges simultaneously by leveraging correct-by-construction synthesis to include confidentiality and availability. 


Security-related attacks are increasingly becoming pervasive in safety-critical CPS. While most of the well-known attacks---such as drone hacking \cite{droneHack16}, Jeep hacking \cite{JeepHack15}, pacemaker and Implantable Cardioverter Defibrillator (ICD) attacks \cite{halperin2008pacemakers,HijackInsulin11}---exploit unencrypted wireless communication, such attacks can be readily guarded against by following recommended cryptographic measures without requiring any significant modification to the control logic. On the other hand, security vulnerabilities related to information leaks via side-channels may be impossible to mitigate without requiring a non-trivial modification to control software, as the side-channels are products of the interaction of the embedded control software with its physical environment.


To provide a simple scenario of unintended information leak via timing side-channels, let us consider an example in the setting of smart hospitals shown in Figure~\ref{fig:smart-hospital}. An increasing prevalence of smart-devices and sensors in modern hospitals makes such an attack scenario on smart hospitals viable. While at a first glance, this example may seem contrived, it emphasizes how seemingly innocuous observations can provide a strong side-channel to leak private information. Furthermore, the presence of wide variety of observations (time delays between various responses \cite{Leu18}, temperature \cite{Hutter13}, electro-magnetic emissions \cite{Mai2012}, optical \cite{Mai2012} and acoustic \cite{genkin}, physiological \cite{Mohse16}) in CPS expose corresponding attack surfaces to the intruder and render CPS even more vulnerable than traditional software.



\begin{figure*}[t]
  \centering
  	\subfigure[]{
	\includegraphics[height=0.19\textwidth]{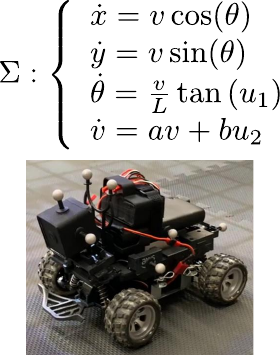}
	}\label{fig:a}
  	\subfigure[]{
  		\includegraphics[height=0.17\textwidth]{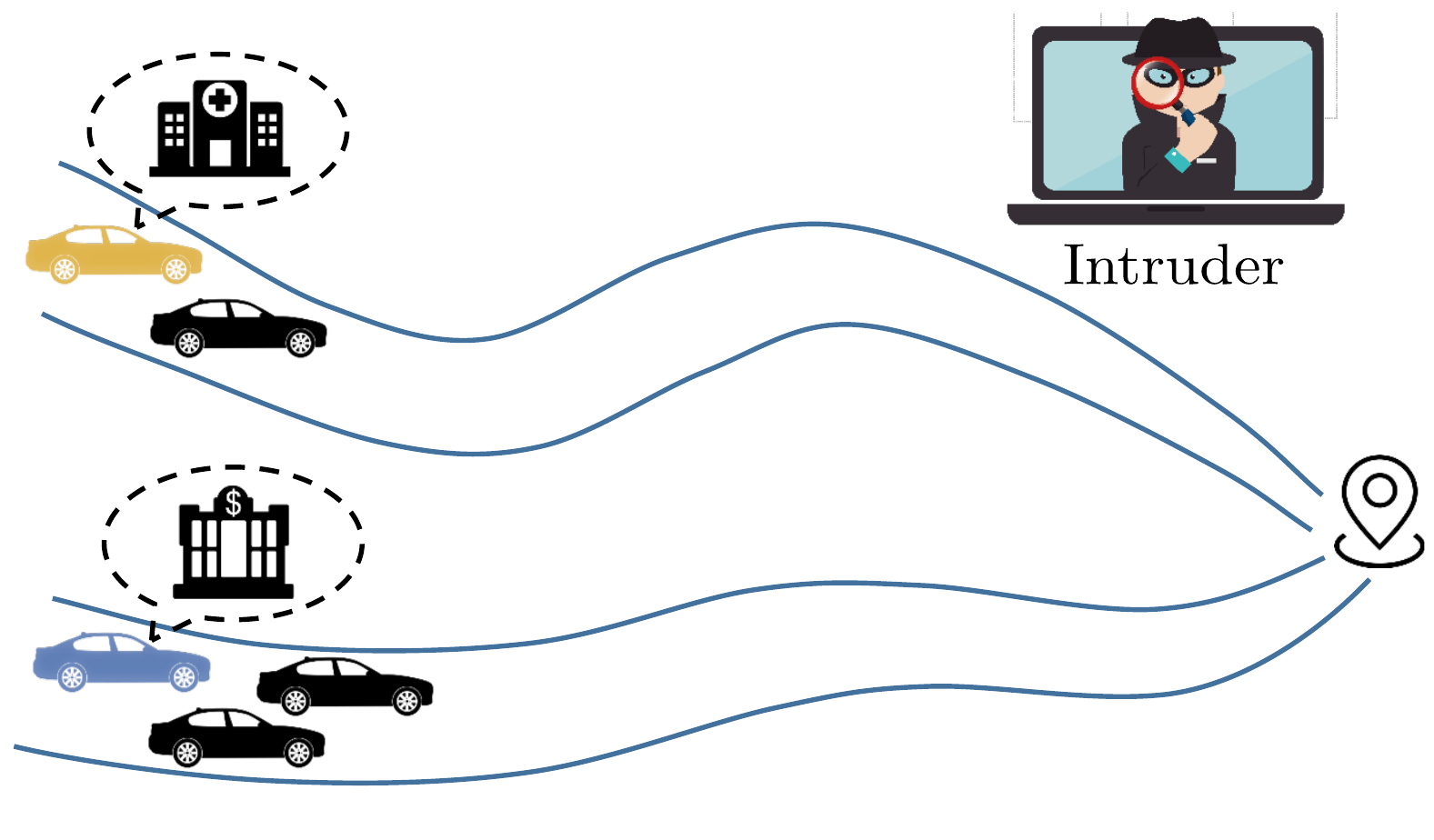}
 }\label{fig:b}
		\subfigure[]{
	\includegraphics[height=0.19\textwidth]{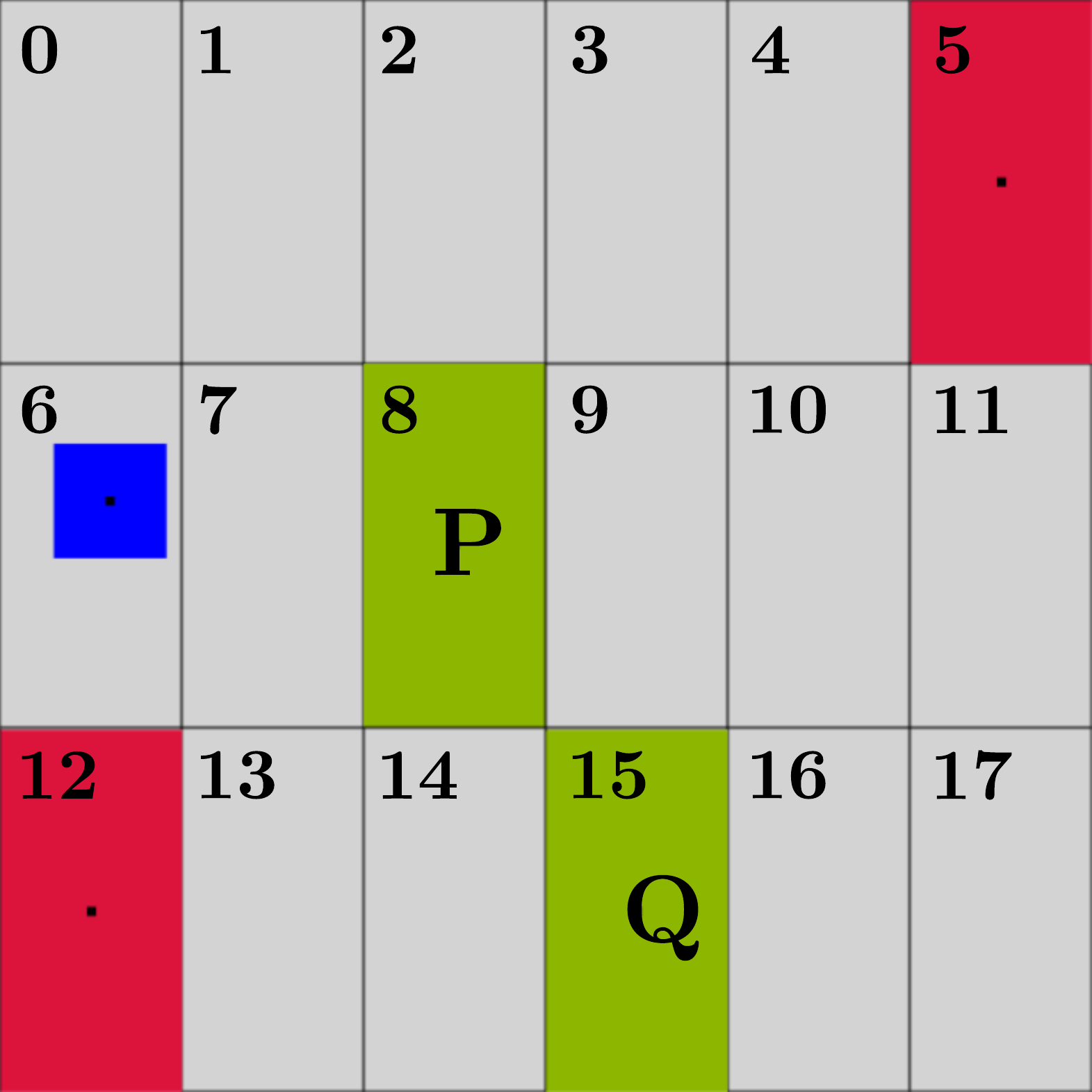}} \label{fig:c}
	 \subfigure[]{
	\includegraphics[height=0.19\textwidth]{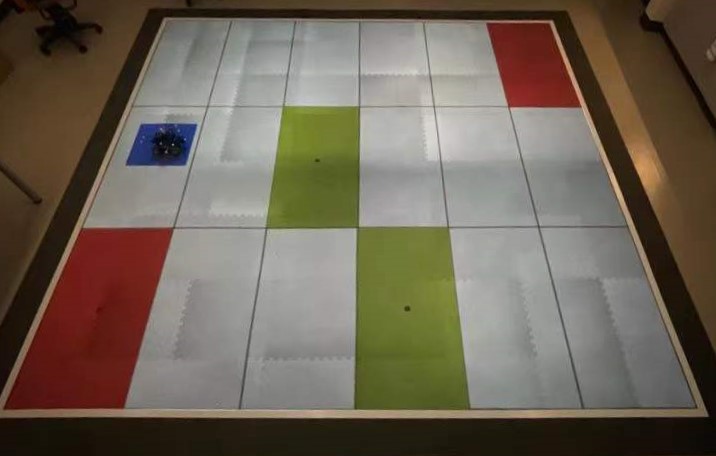}}
          \caption{\small The AWS DeepRacer car and its dynamics (a); The plausible deniability of the car for secret initial region (b);
          The grid-world observations (c) where the red regions depict sensitive starting locations (e.g., hospital or bank) and the green regions represent the target; Our actual platform in the lab (d) corresponding to this grid-world.} \label{platform}
        \vspace{-0.3cm}
\end{figure*}
Formal-methods based approach to system design \cite{tab09,belta}  recommends rigorous requirement specification in every stage of the system development. Formal verification \cite{katoen08} and controller synthesis \cite{tab09,belta} are two leading approaches to provide correctness guarantees with respect to such requirements. While formal verification aims at providing a proof of correctness with respect to the given specifications, the goal of the controller synthesis approach is more ambitious: it takes a control system together with the specification, and produces a controller such that the resulting closed-loop satisfies the specification. The automated controller synthesis approach from formal requirements is referred to as correct-by-construction controller synthesis scheme  \cite{tab09,belta,lee2}. While the controller synthesis approach has been well understood for safety, the secrecy requirements in CPS are often verified after the design of controllers. Hence, if the system leaks information, the controller needs to be redesigned incurring high verification and validation costs.

We envisage a paradigm shift in the development of simultaneously safe and secure CPS that advocates a \textbf{secure-by-construction} controller synthesis scheme which generalizes existing correct-by-construction synthesis methods by considering privacy properties simultaneously to safety ones during the design phase.

\paragraph{\emph{Overview}}
We give a brief overview of the secure-by-construction approach using a concrete synthesis problem for our experimental setup. 
Consider a physical platform developed as shown in Figure.~\ref{platform}(d).
Here we are interested in synthesizing a controller for the movement of a robotic vehicle (AWS DeepRacer Car in Figure.~\ref{platform}(a)) with safety and security requirements. 
The intuition behind the security property of interest is as follows.
Suppose the initial locations of the vehicle contain critical information which is needed to be kept secret, e.g., the vehicle might be a cash transit van that aims at transferring money initially from a bank to an ATM machine, or a patient who initially visited a hospital but unwilling to reveal this information to others. It is implicitly assumed that there is a malicious intruder who is observing the behavior of the vehicle remotely intending to carry out an attack. Therefore, it is in the interest of the system to verify whether it maintains plausible deniability for secret initial location where some confidential assignment is executed. 
In the physical platform, we assume that the vehicle can start from any of the four corner cell (Cells $0$, $5$, $12$, and $17$). We also assume that Cell 5 and Cell 12  marked in red are sensitive starting locations.
We also assume that the time it takes for the robot to travel to any neighboring cell on east (E), west (W), north (N), and south (S) is the same and it is known to the intruder.  Now assume that the intruder can only observe when the robotic vehicle is in the regions marked by P (parking area) and Q (checkout queue) and gets the common observation G for the rest of the cells. A secure-by-construction controller synthesis task is to design a feedback controller satisfying the following requirements: 1) a mission requirement: the robotic vehicle visits regions $P$ and $Q$ infinitely often and 2) a privacy requirement: the intruder is unable to infer whether the vehicle got initiated from a sensitive location.


Suppose we design a controller providing control strategies from all initial cells such that the robot first follows a shortest path to reach Cell 8 or Cell 15, and then cycles between them forever. It is easy to verify that these control strategies satisfy the mission requirement of visiting regions $P$ and $Q$ infinitely often. However, unfortunately such controller does not satisfy the privacy requirement as it is clear from the following system executions adhering to the aforementioned control strategies: here on the left side we show the system executions, while on the right hand side we show the observations made by the intruder. The notation $\omega$ over parentheses shows the infinite repetition of the finite execution inside them.
\begin{itemize}
    \item  $ 0 {\xrightarrow[]{E}} 1 {\xrightarrow[]{E}} 2 {\xrightarrow[]{S}} 8
  ({\xrightarrow[]{S}} 14 {\xrightarrow[]{E}} 15 {\xrightarrow[]{N}} 9
  {\xrightarrow[]{W}} 8)^\omega 
\qquad  \mapsto 
 G {\xrightarrow[]{}} G {\xrightarrow[]{}} G {\xrightarrow[]{}} P
  ({\xrightarrow[]{}} G {\xrightarrow[]{}} Q {\xrightarrow[]{}} G
  {\xrightarrow[]{}} P)^\omega $
  \item $12 {\xrightarrow[]{E}} 13 {\xrightarrow[]{E}} 14 {\xrightarrow[]{N}} 8
  ({\xrightarrow[]{S}} 14 {\xrightarrow[]{E}} 15 {\xrightarrow[]{N}} 9
  {\xrightarrow[]{W}} 8)^\omega    
  \qquad \mapsto  
  G {\xrightarrow[]{}} G {\xrightarrow[]{}} G {\xrightarrow[]{}} P
  ({\xrightarrow[]{}} G {\xrightarrow[]{}} Q {\xrightarrow[]{}} G
  {\xrightarrow[]{}} P)^\omega $
  \item $5 {\xrightarrow[]{W}} 4 {\xrightarrow[]{W}} 3 {\xrightarrow[]{W}} 2
  {\xrightarrow[]{S}} 8 ({\xrightarrow[]{S}} 14 {\xrightarrow[]{E}} 15
  {\xrightarrow[]{N}} 9 {\xrightarrow[]{W}} 8)^\omega   
\qquad \mapsto  
  G {\xrightarrow[]{}} G {\xrightarrow[]{}} G {\xrightarrow[]{}} G
  {\xrightarrow[]{}} P ({\xrightarrow[]{}} G {\xrightarrow[]{}} Q
  {\xrightarrow[]{}} G {\xrightarrow[]{}} P)^\omega $
  \item $17 {\xrightarrow[]{W}} 16 {\xrightarrow[]{W}} 15 ({\xrightarrow[]{N}} 9
  {\xrightarrow[]{W}} 8 {\xrightarrow[]{S}} 14 {\xrightarrow[]{E}} 15)^\omega 
 \qquad \mapsto  
  G {\xrightarrow[]{}} G {\xrightarrow[]{}} Q ({\xrightarrow[]{}} G
  {\xrightarrow[]{}} P {\xrightarrow[]{}} G {\xrightarrow[]{}} Q)^\omega$
\end{itemize}
  

For this controller, if the system starts in the secret state $12$, the corresponding observation is also matched by the non-secret state $0$. On the other hand, when the system starts in secret state $5$, there is no other non-secret initial state giving the same observation. Hence, whenever the system starts from the secret state $5$, the observation uniquely identifies the initial state to be a secret one. For this controller, we say that the system is not opaque. On the other hand, by modifying the controller to change the strategy from Cell $17$ to the one below makes the system opaque since it matches the observation sequence starting from Cell $5$.

 \begin{itemize}
     \item 
 $17 {\xrightarrow[]{N}} 11 {\xrightarrow[]{W}} 10 {\xrightarrow[]{W}} 9
  {\xrightarrow[]{W}} 8 ({\xrightarrow[]{S}} 14 {\xrightarrow[]{E}} 15
  {\xrightarrow[]{N}} 9 {\xrightarrow[]{W}} 8)^\omega 
 \qquad  \mapsto
  G {\xrightarrow[]{}} G {\xrightarrow[]{}} G {\xrightarrow[]{}} G
  {\xrightarrow[]{}} P ({\xrightarrow[]{}} G {\xrightarrow[]{}} Q
  {\xrightarrow[]{}} G {\xrightarrow[]{}} P)^\omega$
 \end{itemize}

We detail the secure-by-construction synthesis framework to \textbf{automatically} design such controllers for large-scale CPS satisfying both the complex logic missions as well as the security requirements. 

\begin{figure}[t!]
  \centering
  \includegraphics[width=0.45\textwidth]{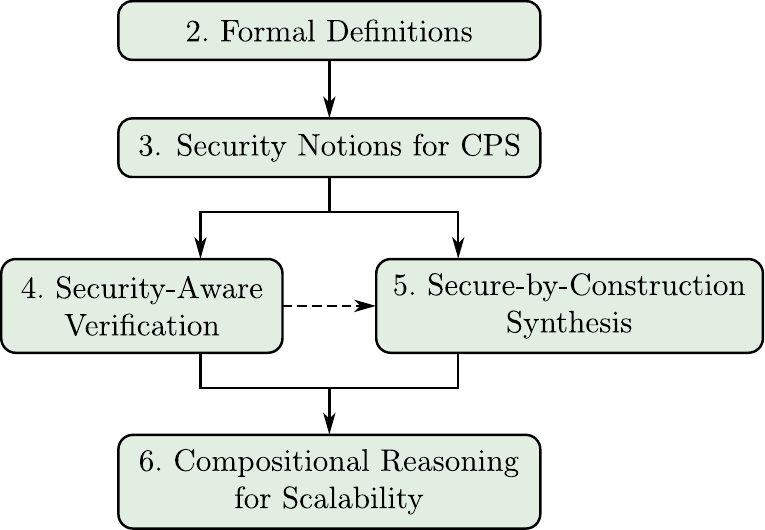}
  \caption{Organization of the paper.}\label{fig:structure}
  \vspace{-0.3cm}
\end{figure}

\paragraph{\emph{Scope}}
The goal of this paper is to provide the reader with a bird-eye view of the recent research and future challenges in this promising and active field. 
We will provide a general definition of the system and provide various definitions from the discrete-event systems (DES), cyber-physical systems (CPS), formal methods (FM) communities. In our selection, the focus of the DES community is on the finite state models, the CS community primarily on the continuous space models, while the results from FM community will primarily focus on logical and automata-theoretic results. We will provide a unifying view of various models and problems studied in this context, and then survey key complexity and (un-) decidability results  while providing practical sub-classes and theoretical tools studied to recover efficient solutions.
A particularly fruitful avenue to provide scalability is \emph{compositional reasoning} and we will present a separate treatment on compositional verification and synthesis. 
The organization of the paper is graphically depicted in Figure~\ref{fig:structure}.  

\section{Preliminaries}\label{sec: pre}
\paragraph{\emph{Notation}} 
We denote by $\R$ and $\N$ the set of real numbers and non-negative integers, respectively. These symbols are annotated with subscripts to restrict them in
the usual way. We use notations $\mathcal{K}$ and $\mathcal{K}_{\infty}$ to denote the different classes of comparison functions, as follows: $\mathcal{K} = \{\gamma:\mathbb R_{\ge 0}\rightarrow\mathbb R_{\ge 0}   :  \gamma \text{ is continuous, strictly increasing and }$ $ \gamma(0)=0\}$; $\mathcal{K}_{\infty} = \{\gamma \in \mathcal{K}   : \lim_{r\rightarrow \infty}\gamma(r) =\infty\}$.
Given $N\in\N_{\ge1}$ vectors $\nu_i\in\R^{n_i}$, $n_i\in\N_{\ge1}$, and $i\in[1;N]$, we
write $\nu=(\nu_1,\ldots,\nu_N)$ to denote the corresponding concatenated vector in $\R^n$ with
$n=\sum_i n_i$.
Given a vector $x \in \mathbb R^{n}$, we denote the infinity norm of $x$ by $\Vert x\Vert$. We denote by $\mathrm{id}$ the identity function over $\mathbb{R}$, i.e., $\mathrm{id}(r) = r$ for all $r \in \mathbb{R}$. 
The complement of set $X$ w.r.t. $Y$ is defined as $Y \backslash X = \{x : x \in Y, x \notin X\}.$
For any set \mbox{$S\subseteq\R^n$} of the form of finite union of boxes, e.g., $S=\bigcup_{j=1}^MS_j$ for some $M\in\N$, where $S_j=\prod_{i=1}^{n} [c_i^j,d_i^j]\subseteq \R^{n}$ with $c^j_i<d^j_i$, we define $\emph{span}(S)=\min_{j=1,\ldots,M}\eta_{S_j}$ and $\eta_{S_j}=\min\{|d_1^j-c_1^j|,\ldots,|d_{n}^j-c_{n}^j|\}$. Moreover, for a set in the form of $X= \prod_{i=1}^N X_i$, where $X_i \subseteq \R^{n_i}$ are of the form of finite union of boxes, and any positive (component-wise) vector $\eta = (\eta_1,\dots,\eta_N)$ with $\eta_i \leq \emph{span}(X_i)$, $\forall i\in [1;N]$, we define $[X]_\eta= \prod_{i=1}^N [X_i]_{\eta_i}$, where $[X_i]_{\eta_i} = [\R^{n_i}]_{\eta_i}\cap{X_i}$ and  $[\R^{n_i}]_{\eta_i}=\{a\in \R^{n_i}: a_{j}=k_{j}\eta_i,k_{j}\in\mathbb{Z},j=1,\ldots,n_i\}$.

For a set $A$, we write $A^*$ for the set of finite sequences from $A$ and $A^\omega$ for the set of (infinite) $\omega$-sequences. We write $A^\infty = A^* \cup A^\omega$.  


\begin{tcolorbox}
\begin{definition}\upshape(\textbf{System Model}) 
A system $\Sigma$ in this paper is described by a quadruple
\begin{equation}
\Sigma=(X,X_0,U,\rTo),
\end{equation}
where
$X$ is a (possibly infinite) set of states, 
$X_0\subseteq X$ is a (possibly infinite) set of initial states,
$U$ is a (possibly infinite) set of inputs, and
$\rTo\subseteq X\times U\times X$ is a transition relation.
We call a system \textit{finite} (or \textit{symbolic}), if $X$ and $U$ are finite sets. \end{definition}
\end{tcolorbox} 
A transition \mbox{$(x,u,x')\in\rTo$} is also denoted by $x\rTo^ux'$.
For a transition $x\rTo^ux'$, state $x'$ is called a \mbox{$u$-successor}, or simply a successor, of state $x$; state $x$ is called a \mbox{$u$-predecessor}, or simply a predecessor, of state $x'$. We  denote by $\mathbf{Post}_{u}(x)$ the set of all \mbox{$u$-successors} of state $x$
and by $\mathbf{Pre}_{u}(x)$ the set of all \mbox{$u$-predecessors} of state $x$.
For a set of states $q\in 2^X$, we write
\[
    \mathbf{Post}_{u}(q)=\cup_{x\in q}\mathbf{Post}_{u}(x) \text{  and  }   \mathbf{Pre}_{u}(q)=\cup_{x\in q}\mathbf{Pre}_{u}(x).
\]
We call a system \textit{deterministic}, if for any state $x\in{X}$ and any input $u\in{U}$, $\mathbf{Post}_{u}(x)$ is singleton; otherwise we call it \textit{non-deterministic}.

A system $\Sigma$ from an initial state $x_0\in X_0$ and input sequence $u_1u_2\cdots u_n\in U^*$, induces a finite state \emph{run} 
\begin{align}\label{run}
x_0\rTo^{u_1}x_1\rTo^{u_2}\cdots\rTo^{u_{n-1}}x_{n-1}\rTo^{u_{n}}x_n,
\end{align}
such that $x_i\rTo^{u_{i+1}}x_{i+1}$ for all $0\leq i<n$. 
Note that the run induced by an input sequence may not be unique because the system may be non-deterministic. 

We call a finite sequence of states $x_0x_1\cdots x_n\in X^*$ a \emph{finite path} of the system $\Sigma$ and denote by $\texttt{Path}(\Sigma,x_0)$ the set of all finite paths generated by $\Sigma$ starting from $x_0$ with $\texttt{Path}(\Sigma)=\cup_{x_0\in X_0}\texttt{Path}(\Sigma,x_0)$. 
Similarly, an infinite path $x_0x_1\dots \in X^\omega$ is an $\omega$-sequence defined analogously and we denote by
$\texttt{Path}^\omega(\Sigma,x_0)$ the set of all infinite paths of $\Sigma$ from $x_0$ with $\texttt{Path}^\omega(\Sigma)=\cup_{x_0\in X_0}\texttt{Path}^\omega(\Sigma,x_0)$. 
 
\paragraph{\emph{Behaviors}} 
A primary concern is whether the behaviors of system $\Sigma$ satisfy some desired specification. 
Formally, let $\mathcal{AP}$ be a finite set of features, or \emph{(atomic) propositions}, of the state space.
We view the states with the lenses of atomic propositions, and to do so, we define a \emph{labeling function} $L: X\to 2^\mathcal{AP}$
that assigns to each state $x\in X$ in $\Sigma$ a set of propositions $L(x)$ true at the state $x$.  
The labeling function can naturally be extended from states to path: we call such labeling of a path a \emph{trace}. 
For any finite or infinite path $\mathbf{x}=x_0x_1\cdots \in X^\infty$, its trace is  $L(\mathbf{x})=L(x_0)L(x_1)\cdots \in (2^{\mathcal{AP}})^\infty$. 
The set of all  finite traces and the set of all infinite traces are denoted by $\texttt{Trace}(\Sigma)$ and $\texttt{Trace}^\omega(\Sigma)$, respectively.

\paragraph{\emph{Observations}} 
The system releases information to the external world during its execution. 
The external world often may not observe the internal states $X$ or their atomic propositions directly but rather their properties over some observation symbols.
Let $Y$ be such set of observations. 
Let the \emph{output function} $H: X\to Y$
determine the external observation of each internal state $x\in X$. 
It can naturally be extended to finite or infinite paths, i.e., for a path $\mathbf{x}=x_0x_1\cdots  \in X^\infty$, 
its \emph{output} corresponds to a sequence 
$H(\mathbf{x})=H(x_0)H(x_1)\cdots \in Y^\infty$. 

The system $\Sigma$ is said to be \textit{metric} if the observation set $Y$ is equipped with a metric
$\mathbf{d}:Y\times Y\rightarrow\mathbb R_{\ge 0}$. 
For any two paths $\mathbf{x}=x_0x_1\cdots $ and $\mathbf{x}'=x_0'x_1'\cdots $, 
we say the outputs of  $\mathbf{x}$ and $\mathbf{x}'$ are 
\emph{(exactly) output equivalent}, denoted by  $H(\mathbf{x})=H(\mathbf{x}')$, if  $H(x_i)=H(x_i')$ for all $i \geq 0$; 
on the other hand, we say that they are
\emph{$\delta$-approximately output equivalent}, and write $H(\mathbf{x}){\approx_\delta} H(\mathbf{x}')$, if   $\sup_{i\geq 0} \mathbf{d}( H(x_i),H(x_i'))\leq \delta$.

To emphasize the labeling $L: X\to 2^\mathcal{AP}$ and output functions $H: X\to Y$ of a system $\Sigma$, we rewrite the tuple describing the system as 
\[
\Sigma=(X,X_0,U,\rTo,\mathcal{AP}, L, Y, H).
\]
When it is clear from the context, we may drop some of the elements in the tuple for the sake of simple presentation.
\begin{remark}
In the DES literature, it is customary to model a system as a finite state machine
$G=(X,E,\delta,X_0)$, where $X$ is a set of states, $E$ is a  set of events, $\delta:X\times E \to 2^X$ is a transition function and $X_0\subseteq X$ is a set of initial states  \cite{cassandras2021introduction}.  
In such treatments, both inputs and properties are captured by events $E$.  Furthermore, it is also assumed that the observation mapping is also event-based captured by  a natural projection $P:E\to E_o$.   
\end{remark}
Our modeling framework is general enough to capture treatment in DES literature and capable of expressing more general scenarios posed in the reactive control systems settings.

\section{Security of CPS}
Security requirements, in the DES \cite{lin2011opacity,Wu2013ComparativeAnalysisOpacity,yin2017AutoTW,lafortune2018history} and control theory communities,
are often expressed using the notion of opacity, while in the realm of computer science security
requirements are expressed using closely related, but subtly different, concepts of non-interference \cite{milushev2012noninterference,DBLP:conf/icse/nilizadeh,wu2018eliminating}, K-safety \cite{sousa2016cartesian,pasareanu2016multi}, language-based secrecy \cite{ACZ06}, and their generalizations using HyperLTL properties \cite{Clark10,Clark14}.
We review these notions in this section.


\subsection{Security Notions for Finite Systems: Opacity}

\

\begin{tcolorbox} 
\textbf{Attack Model.} In the setting discussed here, 
we assume that there exists a  \emph{secret predicate}  on runs that models the confidential behavior of the system.
 The system does not want the intruder to infer the status of the
 secret predicate, i.e., whether it has executed a secret run.
  We consider that the intruder knows the dynamics of the system; and can 
  observe the output sequences of the system.  
  The intruder wants to use the output sequences observed online and the knowledge of the system model to infer 
  certain information about the secret predicates of the corresponding run.  For simplicity, we assume that the input sequences are internal information and  unknown to the intruder.
  This setting can be easily relaxed to handle the case where both input and output information are available to the intruder.
\end{tcolorbox}

\emph{Opacity} is a well-studied confidentiality property that captures whether or not the ``secret" of the system can be revealed to an intruder that can infer the system's actual behavior based on the information flow.
A system is said to be opaque if it always has the plausible deniability for any of its secret behavior.

\begin{tcolorbox} 
\begin{definition}[\textbf{Language-Based Opacity}]\label{def:opa-finite}\upshape 

For a system $\Sigma=(X,X_0,U,\rTo, Y, H)$, let $\mathcal{P}_S\subseteq \texttt{Path}(\Sigma)$ be the set of secret (finite) paths and $\mathcal{P}_{P}\subseteq \texttt{Path}(\Sigma)$ be a set of non-secret  (finite) paths.
We say system $\Sigma$ is \textbf{opaque} w.r.t.\ $\mathcal{P}_S$ and $\mathcal{P}_{P}$ if 
for any secret path $\mathbf{x}\in  \mathcal{P}_{S}$, there exists a non-secret path $\mathbf{x}'\in \mathcal{P}_{P}$ such that  $H(\mathbf{x})=H(\mathbf{x}')$. 
\end{definition}
\end{tcolorbox}

The above definition of opacity is referred to as \emph{language-based opacity} in the DES literature  \cite{lin2011opacity} as it uses languages $\mathcal{P}_S$ and $\mathcal{P}_{P}$ to represent secret and non-secret behaviors, respectively. 
The condition in the definition can also be equivalently written in terms of language inclusion as follows: 
\begin{equation} 
 H(  \mathcal{P}_{S}) \subseteq   H(  \mathcal{P}_{P}). 
\end{equation}

In specific applications,   secret paths $\mathcal{P}_S$ usually have concrete meanings, e.g., currently at a secret location or initiated from a secret location. 
Therefore, a commonly used approach is to consider a set of \emph{secret states} $X_S\subseteq X$.   
Depending on what information the system wants to hide, the following \emph{state-based} notions of opacity have been introduced in the literature. 
 
\begin{tcolorbox}[breakable, enhanced] 
\begin{definition}[\textbf{State-Based Opacity}]\label{def:opa-state}\upshape 
Let $\Sigma=(X,X_0,U,\rTo,Y,H)$ be a  system, $X_S\subseteq X$ be a set of secret states and $K\in \mathbb{N}$ be a non-negative integer.
We say system $\Sigma$ is  
\begin{itemize}[leftmargin=*]
  \item
  \emph{Initial-State Opaque} \cite{saboori2013verification} if for any  path  $\mathbf{x}=x_0x_1\cdots x_n\in \texttt{Path}(\Sigma)$, where $x_0\in X_S$, 
  there exists a path   $\mathbf{x}'=x_0'x_1'\cdots x_n'\in \texttt{Path}(\Sigma)$, where  $x_0'\notin  X_S$, 
  such that $H(\mathbf{x})=H(\mathbf{x}')$;  
  \item
  \emph{Current-State Opaque} \cite{saboori2007notions} if for any path  $\mathbf{x}=x_0x_1\cdots x_n\in \texttt{Path}(\Sigma)$, where $x_n\in X_S$,
  there exists a path $\mathbf{x}'=x_0'x_1'\cdots x_n'\in \texttt{Path}(\Sigma)$, where $x_n'\notin X_S$,  such that  $H(\mathbf{x})=H(\mathbf{x}')$;  
  \item
  \emph{Infinite-Step Opaque} \cite{saboori2012verification} if  for any path $\mathbf{x}=x_0x_1\cdots x_n \dots x_{n+k}\in \texttt{Path}(\Sigma)$, where $x_n{\in} X_S$,
  there is a path $\mathbf{x}'=x_0'x_1'\cdots x_n' \dots x_{n+k}'\in \texttt{Path}(\Sigma)$, where $x_n'\notin  X_S$, such that  $H(\mathbf{x}){=}H(\mathbf{x}')$; 
  \item
  \emph{$K$-Step Opaque} \cite{saboori2011verificationk} if for any  path  $\mathbf{x}=x_0x_1\cdots x_n \dots x_{n+k}\in \texttt{Path}(\Sigma)$, where $x_n\in X_S$ and $k\leq K$,
  there exists a path  $\mathbf{x}'=x_0'x_1'\cdots x_n' \cdots x_{n+k}'\in \texttt{Path}(\Sigma)$, where $x_n'\notin  X_S$, such that  $H(\mathbf{x})=H(\mathbf{x}')$;  
  \item 
  \emph{Pre-Opaque} \cite{yang2020pre} if   
  for any  path $\mathbf{x}=x_0x_1\cdots x_n$ and  any $k\in \mathbb{N}$, 
  there exists a path  $\mathbf{x}'=x_0'x_1'\cdots x_n'  \cdots x_{n+k}'  \in \texttt{Path}(\Sigma)$, where   $x_{n+k}'\notin   X_S$, such that   $H(x_0x_1\dots x_n)=H(x_0'x_1'\cdots x_n')$. 
\end{itemize}
\end{definition}
\end{tcolorbox} 

\begin{figure}[t]
  \centering
  \includegraphics[width=0.45\textwidth]{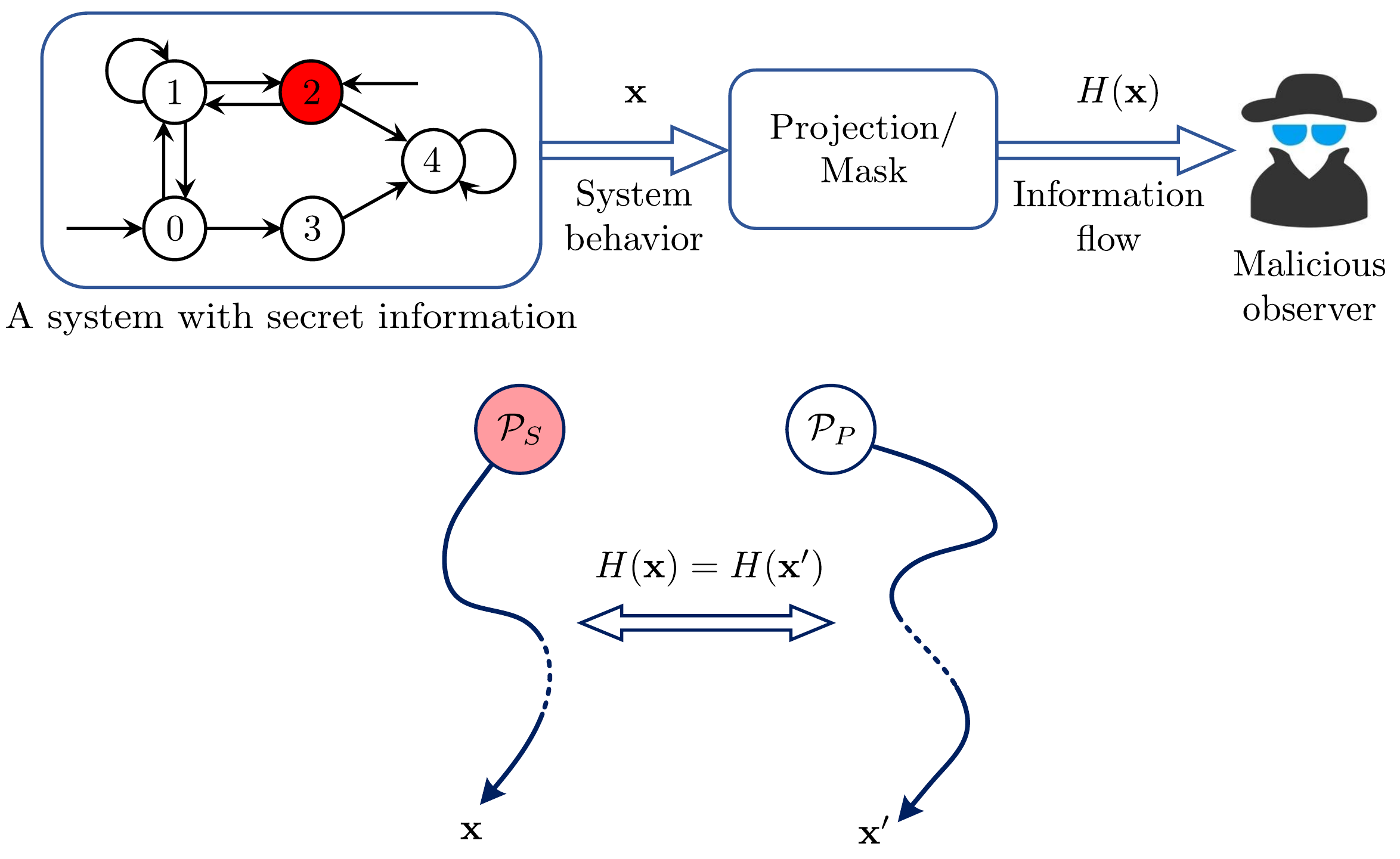}
  \caption{Initial-state opacity.}\label{fig:opacity}
  \vspace{-0.3cm}
\end{figure}

The above state-based notions of opacity are closely related to the three fundamental state estimation problems in the systems theory: filtering, smoothing and prediction \cite{hadjicostis2020estimation}. 
Specifically, current-state opacity is  related to the \emph{filtering} problem because it requires that the intruder can never determine for sure that the system is currently at a secret state. 
Initial-state opacity and  infinite/$K$-step opacity are related to the \emph{smoothing} problem because they both consider the scenario where the intruder can use latter observations to infer whether or not a system was at a secret state for some previous or the initial instant. 
In particular, initial-state opacity says that the intruder can never know that the system was initiated from a secret state, 
and $K$-step opacity says that the intruder can never know that the system was at a secret state within the past $K$-steps. 
Clearly, when $K$ takes values $0$ and $\infty$, $K$-step opacity becomes current-state opacity and infinite-step opacity, respectively. 
Finally, the notion of pre-opacity is related to the \emph{predication} problem by requiring that the intruder can never know for sure that the system will reach a secret state for some specific future instant. This type of opacity essentially captures the intention security of the system. An illustration of the concept of initial-state opacity is depicted in Figure.~\ref{fig:opacity}.

\subsection{Security Notions for CPS: Approximate Opacity}



The formulation of opacity in the last subsection requires that for any secret behavior, there exists a non-secret behavior
such that they generate exactly the same output. Therefore, we will also refer to these definitions as \emph{exact opacity}.
Exact opacity essentially assumes that the intruder or the observer can always measure each output or distinguish
between two different outputs precisely. This setting is reasonable for non-metric systems where outputs are
symbols or events. However, for metric systems, e.g., when the outputs are physical signals, this setting is too
restrictive. In particular, due to the imperfect measurement precision, which is almost the case for all physical
systems, it is very difficult to distinguish two observations if their difference is very small. Therefore, exact
opacity may be too strong for metric systems and it is meaningful to define a weak and ``robust" version of opacity.  

In \cite{Yinapproximate},  a concept called \emph{approximate opacity} is proposed  that is  more applicable to metric systems. 
The new concept can be seen as a ``robust" version of opacity by characterizing under what measurement precision the system is opaque.
In particular, we treat two outputs as ``indistinguishable" outputs if their distance is smaller than a given threshold parameter $\delta\geq 0$, 
i.e., condition $H(\mathbf{x})=H(\mathbf{x}')$ is replaced by $H(\mathbf{x})\approx_{\delta}H(\mathbf{x}')$. 
All exact notions of opacity defined in Definition~\ref{def:opa-state} can be generalized to the approximate versions by replacing the output equivalence condition as $\delta$-closeness. In the remainder part of this paper, for the sake of simple presentation, we mainly focus on initial-state opacity to present the main results. 
Moreover, when discussing state-based opacity, we incorporate the secrete state set $X_S$ in the system definition and use $\Sigma=(X,X_0,X_S,U,\rTo,Y,H)$ to denote a metric system.

\begin{tcolorbox} 
\begin{definition}[\textbf{Approximate Opacity}]\label{def:opa-app}\upshape 
Let $\Sigma=(X,X_0,X_S,U,\rTo,Y,H)$ be a metric system, with the metric $\mathbf{d}$ defined over the output set, and a constant $\delta\geq 0$.
System $\Sigma$ is said to be 
  \emph{$\delta$-approximate initial-state opaque} if for any  path  $\mathbf{x}=x_0x_1\cdots x_n\in \texttt{Path}(\Sigma)$, where $x_0\in X_S$,  
  there exists path     $\mathbf{x}'=x_0'x_1'\cdots x_n'\in \texttt{Path}(\Sigma)$, where  $x_0'\notin X_S$,  
  such that  $H(\mathbf{x})\approx_{\delta}H(\mathbf{x}')$.
\end{definition} 
\end{tcolorbox} 
 
Clearly, when $\delta=0$,  $\delta$-approximate initial-state    opacity reduces to its exact version  in Definition~\ref{def:opa-state}.
The main difference is how we treat two outputs as indistinguishable outputs. 
Specifically, same as in the exact case, we still assume that the intruder know the system model and the output trajectory generated. 
However, we further assume that the intruder may not be able to distinguish an output trajectory from other $\delta$-closed ones.
Intuitively, the approximate version of opacity can be interpreted as
``\emph{the secret of the system cannot be revealed to an intruder that does not have an enough measurement precision related to parameter $\delta$}".
In other words, instead of providing an exact security guarantee, approximate opacity provides a relaxed and quantitative security guarantee with respect to the measurement precision of the intruder. 
Therefore, the value $\delta$ can be interpreted as either the measurement imprecision of the intruder 
or the security level the system can guarantee, i.e., under how powerful intruder the system is still secure.

\subsection{Safety \& Security in Formal Methods: Temporal Logic} 
In the DES literature, opacity is defined over (possibly arbitrarily long) finite paths. 
In the context of formal verification and synthesis in the computer science literature, formal properties are usually defined over infinite traces. 
Specifically, a property $\mathcal{P}\subseteq (2^{\mathcal{AP}})^\omega$ is a subset of infinite traces. 
Since languages over infinite sequences are more expressive than languages over finite ones, it is more general to consider $\omega$-languages than finite-languages. 
Formal logics such as LTL~\cite{katoen08} and their generalizations (hyperLTL~\cite{Clark14,Clark10}) are convenient ways to express subsets of $\omega$-regular languages.

\paragraph{\emph{Safety and Mission Requirements}}
Linear Temporal Logic (LTL) \cite{katoen08} is a convenient and expressive
formalism to express properties of infinite runs (or traces) of the system.
A restricted form of LTL~\cite{DeGia13} has been proposed to express properties
of finite runs or traces. 
The set of LTL properties over the atomic proposition $\mathcal{AP}$ can be defined by the following grammar: 
\[
\phi  ::=  a \in \mathcal{AP} \mid \neg \phi  \mid  \phi \lor \phi  \mid 
\nextt \phi  \mid  \phi \until \phi.
\]
Here, $\neg$ and $\vee$ stand for logical negation and disjunction, while $\nextt$ and $\until$ are temporal modalities expressing \texttt{next} (in the next discrete step) and \texttt{until} (left property continues to hold until the property on the right holds) modalities, respectively.
For convenience, additional operators can be derived from these basic ones:
$\texttt{true} \rmdef a \vee \neg a; \texttt{false} \rmdef \neg \texttt{true}; \varphi \wedge \psi \rmdef \neg (\neg \varphi \vee \neg \psi);
\varphi \rightarrow \psi \rmdef \neg \varphi \vee \psi;
\eventually \varphi \rmdef \texttt{false} \until \varphi;\text{  and }
\always \varphi \rmdef \neg \eventually \neg \varphi$.
Here $\wedge$ and $\rightarrow$ stand for conjunction and implication, while $\eventually$ and
$\always$ stand for temporal operators \texttt{finally} (some time in the future)
and \texttt{globally} (at each step).
The semantics of the LTL can be defined inductively in a straightforward
fashion (see, \cite{katoen08}).
This logic allows the designers to unambiguously
characterize system properties.
For instance, a safety property can be expressed as ``$\always \neg \phi$''
which states that some bad property $\phi$ never holds.
Similarly, a reachability property ``$\eventually \phi$'' can be used to express
that some good property $\phi$ eventually holds.

For an infinite trace $r \in \texttt{Trace}^\omega(\Sigma)$ of a system $\Sigma$, we say that $r$ satisfies the LTL property $\varphi$ and denoted by $r \models \varphi$, if it satisfies the LTL formula $\varphi$.   
It is known that the set of all infinite traces satisfying an LTL formula can be accepted by either a non-deterministic B\"{u}chi automaton or a deterministic Rabin automaton \cite{katoen08}.
Given a system $\Sigma$ and an LTL requirement $\varphi$, we denote by $\Sigma \models \varphi$ if for every infinite trace $r \in \texttt{Trace}^\omega(\Sigma)$ we have that $r \models \varphi$.

LTL formulae capture  the safety and functional correctness requirements of the system. 
Essentially, it evaluates whether or not each single infinite trace satisfies the property. 
However, formal reasoning about security properties requires reasoning with multiple traces of the system.
For example, Alur et al.~\cite{ACZ06} show that modal $\mu$-calculus is insufficient to express all opacity policies.

Clarkson and Schneider~\cite{Clark10} introduced the concept of hyperproperties to express security policies using second-order logic. 
Hyperproperties generalize the concept of linear-time properties~\cite{katoen08} from being sets of runs to {\it sets of sets of runs}.
HyperLTL, unlike LTL which implicitly considers only a single trace at a time, can relate different trace executions simultaneously through
the use of existential and universal quantifiers. 
The HyperLTL formulae can be given using the following grammar: 
\begin{eqnarray*}
\psi &::=& \exists \pi. \psi \mid \forall \pi. \psi \mid  \phi \\
\phi &::=& a_\pi \mid \neg \phi\mid \phi \lor \phi \mid
\nextt \phi \mid \phi \until \phi.
\end{eqnarray*}

The key distinction over LTL formulae is the introduction of trace quantifiers
$\exists$ and $\forall$. The quantifier $\exists \pi$ stands for ``for some
trace $\pi$" while the quantifier $\forall \pi$  stands for ``for all traces
$\pi$", respectively. 
The variable $\phi$ generates standard LTL formulae (complete with Boolean
connectives and temporal operators $\nextt$ and $\until$)  with the exception that
atomic propositions can refer to distinct trace variables.
Hence, for every proposition $a \in \mathcal{AP}$ and trace variable $\pi$, we
use $a_\pi$ to express that proposition $a$ is referring to the trace $\pi$.   
We say that a trace variable occurs free in a HyperLTL formula, if it is not
bounded by any trace quantifier.
A HyperLTL formula with no free variable is called a closed formula. 
 
 HyperLTL can express certain opacity properties.
For instance, the following $\text{HyperLTL}$ formula expresses language-based opacity introduced in
Definition \ref{def:opa-finite} when $\mathcal{P}_S$ and $\mathcal{P}_P$ are given as LTL properties $\varsigma$ and $\varphi$ 
\[
\forall \pi \exists \pi'\cdot    L(\pi) \models \varsigma \to  ( H(\pi) = H(\pi')  \wedge L(\pi')\models \varphi)   
\]
where  $\pi$ is defined over $\texttt{Path}^\omega(\Sigma)$.

Unfortunately, since HyperLTL requires quantification over paths in the
beginning of the formula, it is not expressive enough to define infinite-step, current-state, and $K$-step opacity requirements.

We propose the following generalized language-based opacity notion which extends language-based opacity in Definition~\ref{def:opa-finite} from finite paths to infinite paths.
\begin{tcolorbox} [breakable, enhanced] 
\begin{definition}[\textbf{Generalized Language-Based Opacity}]\label{def:op-general}\upshape 
Let $\Sigma=(X,X_0,U,\rTo, \mathcal{AP}, L, Y, H)$ be a metric system, with the metric $\mathbf{d}$ defined over the output set, and a constant $\delta\geq 0$,
$\mathcal{P}_S\subseteq \texttt{Trace}^\omega(\Sigma)$ be a secret property and $\mathcal{P}_{P}\subseteq \texttt{Trace}^\omega(\Sigma)$ 
be a public property. 
For computational representation, the secret and public properties can be expressed either logically (e.g., via LTL) or using automatic structures (e.g., $\omega$-automata or finite state machines).

We say system $\Sigma$ is \textbf{opaque} with respect to $\mathcal{P}_S$ and $\mathcal{P}_{P}$ if for any secret path $\mathbf{x}\in \texttt{Path}^\omega(\Sigma)$, where $L(\mathbf{x})\in  \mathcal{P}_{S}$, there exists a non-secret path $\mathbf{x}'\in \texttt{Path}^\omega(\Sigma)$, where $L(\mathbf{x}')\in  \mathcal{P}_{P}$, such that  
\[
    H(\mathbf{x})\approx_\delta H(\mathbf{x}').
\]
\end{definition}
\end{tcolorbox} 
The above definition of language-based opacity generalizes  Definition~\ref{def:opa-finite} in threefold. 
First, secret behaviors are defined in terms of traces rather than the internal paths. 
This setting clearly subsumes Definition~\ref{def:opa-finite}  because we can set the labeling function as an identity mapping $L:X\to X$. 
Second, secret behaviors are evaluated in terms of infinite sequences rather than finite sequences.  
Note that,  state-based notions of opacity in Definition~\ref{def:opa-state} are instances of Definition~\ref{def:opa-finite}. 
Therefore, the notions of state-based  opacity, such as initial-state opacity or infinite-step opacity, can all be formulated in terms of Definition~\ref{def:op-general} with a syntactic modification to the system (by adding a dummy sink state to the system) to enable the treatment of finite sequences as infinite sequences.
Finally, Definition~\ref{def:op-general} considers approximate output equivalence rather than the exact one. 
Language-based opacity  in Definition~\ref{def:op-general} also generalizes the notions of \emph{noninterference}~\cite{milushev2012noninterference,DBLP:conf/icse/nilizadeh,wu2018eliminating} and $2$-safety~\cite{sousa2016cartesian,pasareanu2016multi}. 

\paragraph{\emph{Our Settings}} 
In our later problem formulations, for mission/safety requirements we focus on those given as LTL formulae,
while for security ones we focus on generalized language-based opacity in Definition \ref{def:op-general}
where secret and public properties $\alpha_S\subseteq (2^{\mathcal{AP}})^\omega$ and $\alpha_P\subseteq (2^{\mathcal{AP}})^\omega$. 
We denote such a generalized opacity property as a tuple $\alpha = (\alpha_S, \alpha_P)$.
A system $\Sigma$ is called $\alpha$-opaque if it is opaque w.r.t.\ secret and public properties expressed, respectively, using $\alpha_S$ and $\alpha_P$.  
Therefore, we use tuple $(\varphi, \alpha)$ to model both the mission and security requirements. 
Our first objective is to  \emph{verify} whether or not system $\Sigma$ satisfies  $(\varphi, \alpha)$. 
If not, the second objective is to \emph{synthesize} a controller such that the system under control satisfies $(\varphi, \alpha)$. 
We will elaborate in details on the verification and the synthesis problems in Sections~4 and~5, respectively.

\section{Security-Aware Verification}


In the previous section, we have introduced various security formulations that are commonly used from the literature. A natural question to answer is: how to determine whether a given system preserves certain security property? Furthermore, if the system does not preserve the desired security property, how can one design proper controllers to enforce security properties on it? We proceed with the following sections to address these questions. 

In this section, we investigate the  verification problem. 
 \begin{tcolorbox}
\begin{problem}[\textbf{Security-Aware Verification}]\label{prob:verif}\upshape
 Given a mission requirement (as an LTL formula) $\varphi$ and a security property $\alpha$, the security-aware verification problem is to decide whether $\Sigma \models (\varphi, \alpha)$, i.e., $\Sigma$ satisfies the property $\varphi$ and is $\alpha$-opaque.
 \end{problem}
 \end{tcolorbox}
Note that the above problem is formulated in a very general setting  by considering an arbitrary mission requirement $\varphi$ and an arbitrary security requirement $\alpha$.  Throughout the paper, we will mainly consider approximated initial-state opacity as a specified $\alpha$ to present our result.  
To this end,  we first overview the standard model checking approaches for verifying LTL formulae. 
Then, for the verification of security,  we will first discuss the typical schemes on verifying opacity for finite systems, and then present  some recent results which are potential to deal with complex continuous-space CPS.


Given a mission requirement (as an LTL formula) $\varphi$ and a
generalized opacity property (as a pair of two LTL formulae) $\alpha$, the {\it 
verification} problem, $\Sigma \models (\varphi, \alpha)$, can be decomposed into
verifying mission and opacity properties separately. 
The verification problem against the mission requirements given as LTL formula
reduces to a {\it repeated reachability problem} on the composition of
$\Sigma$ with a {\it monitor automaton} corresponding to the negation of the LTL formula~\cite{katoen08}.
The problem is known to be PSPACE-complete and there are efficient symbolic tools (e.g.,
\texttt{NuSMV} \cite{NuSMV} and \texttt{SPIN} \cite{holzmann}) to verify finite labelled transition systems (LTS) representations of $\Sigma$ against LTL requirements.
On the other hand, verification of the generalized language-based opacity has
only been explored in its restricted forms of opacity. We will review them next.


\subsection{Finite Systems} \label{sec:ver}

In the last section, we reviewed a security notion called approximate opacity that is suitable to reason both discrete and continuous dynamics. Here, we show how to verify approximate opacity for finite systems, which will be later used for the verification of opacity for general CPS equipped with continuous state space.
Here we present an approach based on the construction of the   \emph{$\delta$-approximate  observer}.

\begin{tcolorbox}[breakable, enhanced] 
\begin{definition}[\textbf{Approximate Observer}]\label{def:obs}\upshape 
Let $\Sigma=(X,X_0,X_S,U,\rTo,Y,H)$ be a metric system, with the metric $\mathbf{d}$ defined over the output set, and a constant $\delta\geq 0$.
The $\delta$-approximate observer  is a system without outputs
\[
Obs(\Sigma)=(Q,Q_0,U,\rTo_{obs}),
\]
where
\begin{itemize}
  \item
  $Q \subseteq X\times 2^{X\times X}$ is the set of states;
  \item
  $Q_0=\{(x,z)\!\in\! X_0 \times 2^{X_0\times X_0}:   
  (x_I,x_C)\!\in\! z\Leftrightarrow    x_I=x_C\wedge \mathbf{d}(H(x),H(x_C))\leq \delta   \}$ is the set of initial states;
  \item
  $U$ is the set of inputs, which is the same as the one in $\Sigma$;
  \item
  $\rTo_{obs}\subseteq Q\times U\times Q$ is the transition function defined by:
  for any $(x,z),(x',z')\in X\times 2^{X\times X}$ and $u\in U$,   $(x,z)\rTo^{u}_{obs}(x',z')$ if
  \begin{enumerate}
  \item
  $(x,u,x')\in \rTo$; and
  \item
  $z' \!=\! \bigcup_{u'\in U}\bigcup_{(x_I,x_C)\in z} 
  \{ (x_I,x_C'): \mathbf{d}(H(x'),H(x_C'))\!\leq \!\delta   \wedge x_C\rTo^{u'}x_C'  \}$.
  \end{enumerate}
\end{itemize}
For the sake of simplicity, we only consider the part of $Obs(\Sigma)$ that is reachable from initial states.
\end{definition} 
\end{tcolorbox}

Intuitively, the $\delta$-approximate observer works as follows.
Each initial state of $\Sigma$ is a pair consisting of a system state $x\in X_0$ and its $\delta$-closed state pairs $z\in 2^{X_0\times X_0}$. 
Note that each state pair in $z$ is of form $(x_I,x_C)$, where $x_I$ denotes the initial-state the system came from and $x_C$ denotes the current-state of the system.  Note that, since we cannot observe the actual state $x$ precisely, we need to consider all such initial-current state pairs whose second (current-state) component is $\delta$-close to the actual state $x$.   Then from each state, we track states that are consistent with the output information recursively.
Essentially,  the first component can be understood as the ``reference trajectory"  that is used to determine what is ``$\delta$-close" at each instant 
and the second component is the set of ``initial-current-state-pairs" that are $\delta$-close to the reference trajectory. 
This structure is motivated by the well-known ``subset construction" and combines both the initial-state estimator and the current-state estimator in a single structure. 
 
For each state $q=(x,z)\in Q$, we denote by $\texttt{int}(q)=\{ x_I: (x_I,x_C)\in z   \}$ and $\texttt{cur}(q)=\{ x_C: (x_I,x_C)\in z   \}$ 
the set of all possible initial-states and current-states, respectively.   
Employing the above-defined observer, the next theorem is proposed in \cite{Yinapproximate} for the verification of 
$\delta$-approximate initial-state or current-state opacity of finite metric systems.

\begin{tcolorbox}
\begin{theorem}[\textbf{Verification of Opacity}]\label{thm:int} \upshape
Let $\Sigma=(X,X_0,X_S,U,\rTo,Y,H)$ be a finite metric system, with the metric $\mathbf{d}$ defined over the output set, and a constant $\delta\geq 0$.
Let $Obs(\Sigma)=(Q,Q_{0},U,\rTo_{obs})$ be its $\delta$-approximate observer.
Then, $\Sigma$ is $\delta$-approximate initial-state opaque  (respectively, current-state opaque) 
if and only if 
for any $q\in Q$, we have $\texttt{int}(q)\not\subseteq X_S$ (respectively, $\texttt{cur}(q)\not\subseteq X_S$). 
\end{theorem}
\end{tcolorbox}

It is worth noting that the complexity of verifying exact opacity  is already known to be PSPACE-complete \cite{cassez2012synthesis}. 
Therefore, the complexity of verifying approximate opacity is also  PSPACE-complete. 
Essentially, the exponential complexity comes from the subset construction to handle information uncertainty.   
Note that the observer structure presented in Definition~\ref{def:obs} is a unified structure that can handle both initial-state opacity and current-state opacity. 
If one just needs to verify initial-state or current-state opacity, the state space of the observer structure can further be reduced to $X\times 2^X$; see, \cite{Yinapproximate} for more detailed discussion. Regarding the verification of infinite-step or $K$-step opacity, effective algorithms have also been proposed in \cite{saboori2011verificationk,saboori2012verification,yin2017AutoTW} for  the  exact notions and \cite{Yinapproximate} for the approximate ones.   



\subsection{CPS: Abstraction-Based Approach}\label{Sec:veri_abs}

In the previous subsections, we discussed   frameworks on verifying opacity properties for finite systems.
In this subsection, we present some recent results for the verification of opacity for continuous-space CPS based on their \emph{finite abstractions} (a.k.a. symbolic models).

Models of CPS are inherently heterogeneous: from discrete systems modeling computational parts to differential or difference equations modeling continuous physical processes. The ability to handle this heterogeneity is a prerequisite of a rigorous formal framework for both design and analysis framework for CPS. 
In order to address the heterogeneity of CPS models, formal verification and synthesis are often addressed by methods of \emph{abstraction} in which continuous-space
models are approximated by discrete ones. When a suitable finite abstraction is constructed, by leveraging
computational tools developed for DES and games on automata, one can verify or synthesize controllers in an automated fashion against complex
logic requirements. 
\begin{figure}[t!]
  \centering
  \includegraphics[width=0.45\textwidth]{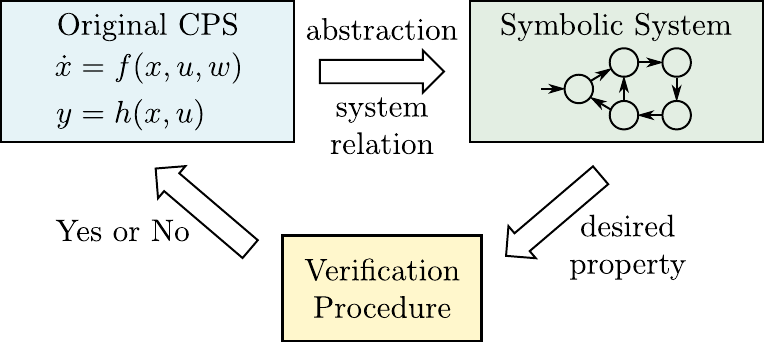}
  \caption{Pipeline of standard discretization-based or abstraction-based verification technique. \vspace{-15pt} }\label{fig:abs_pipeline}
\end{figure}

The pipeline of traditional abstraction-based verification technique is depicted in Figure.\ref{fig:abs_pipeline}, which consists of three key phases. 
The first phase is on the construction of a 
finite abstraction of the CPS with the property
that the set of behaviours of the CPS is included in that of the constructed finite abstraction.
The second phase in the architecture requires symbolic analysis to efficiently reason about formal specifications.
The final phase is to bring the reasoning back to the original concrete systems with formal guarantee.

The key to the construction of such finite/symbolic systems is the establishment of formal relations between the concrete and abstract systems. 
A system relation formalizes the ability to extrapolate properties from an abstraction to the concrete system.
Different system relations enable extrapolation of different kinds of properties.
Such relations include (alternating) (bi)simulation relations, their approximate versions, and strongest or asynchronous $\ell$-complete approximations. 
Finite abstraction together with the notions of so-called simulation relations have been widely and successfully used in the past decade for formal verification, synthesis, and approximation of hybrid systems \cite{AHLP00,girard07,girard2008approximate,girard2,zamani2012symbolic,majid16,pola,tab09,belta,majid8,reissig2017feedback}. 
Nevertheless, none of the constructed finite abstractions in the aforementioned literature is guaranteed to preserve
opacity. As reported in \cite{Zhang2018OpacitySimilation},
existing notions of standard (bi)simulation relations and their approximate versions which are often used in finite abstraction synthesis schemes fail to preserve opacity. 

In the following, we discuss some recent results proposed in \cite{Yinapproximate}, which develop for the first time an abstraction-based opacity verification approach by adapting notions of simulation relations to the context of opacity.

For the sake of an easier presentation, the main results presented in the sequel will be based on the class of discrete-time control systems as follows. 
A discrete-time control system (dt-CS) $\Sigma$ is a metric system and denoted by the tuple $\Sigma\!=\!(X, X_0, X_S, U,f, Y, H)$. 
Notice that here, instead of $\rTo$, we use $f:  X\times  U \rightarrow  X $ to denote the state transition function. The dynamics of $\Sigma$ is described by difference equations of the form
	\begin{align}\label{def:sys1}
	\Sigma:\left\{
	\begin{array}{rl}
	\mathbf{x}(k+1)\!\!\!\!\!\!&=f(\mathbf{x}(k),\nu(k)),\\
	\mathbf{y}(k)\!\!\!\!\!\!&=H(\mathbf{x}(k)),
	\end{array}
	\right.
	\end{align}
where $\mathbf{x}:\mathbb{N}\rightarrow  X $, $\mathbf{y}:\mathbb{N}\rightarrow  Y$, and $\nu:\mathbb{N}\rightarrow  U$ are the state, output, and input signals, respectively.
We write $\mathbf{x}_{x_0,\nu}(k)$ to denote the point reached at time $k$ under the input signal $\nu$ from initial condition $x_0$. Similarly, we denote by $\mathbf{y}_{x_0,\nu}(k)$ the output corresponding to state $\mathbf{x}_{x_0,\nu}(k)$, i.e., $\mathbf{y}_{x_0,\nu}(k)=H(\mathbf{x}_{x_0,\nu}(k))$.

\begin{tcolorbox}[breakable, enhanced] 
\begin{definition}[\textbf{Approximate Initial-State Opacity Preserving Simulation Relation}]\label{InitSOP}\upshape
Consider two metric systems $\Sigma =(X, X_{0},X_{S}, U,f,Y,H)$ and $\hat \Sigma=(\hat X, \hat X_{0},\hat X_{S}, \hat U, \hat f, \hat Y, \hat H)$ with the same output sets $Y = \hat Y$ and metric $\mathbf{d}$.
For $\varepsilon\in \mathbb{R}_{\geq 0}$,
a relation \mbox{$R\subseteq X \times \hat X$} is called an $\varepsilon$-approximate initial-state opacity preserving simulation relation ($\varepsilon$-InitSOP simulation relation) from $\Sigma$ to $\hat \Sigma$
if
\begin{enumerate}[leftmargin=*]
  \item
  \begin{enumerate}
  \item
  $\forall x_{0}\!\in\! X_{0} \cap X_{S},\exists \hat x_{0}\!\in\! \hat X_{0}\cap \hat X_{S}: (x_{0},\hat x_{0})\in R$;
  \item
  $\forall \hat x_{0}\in \hat X_{0}\setminus \hat X_{S},\exists x_{0}\in X_{0}\setminus X_{S}:(x_{0},\hat x_{0})\in R$;
  \end{enumerate}
  \item
  $\forall (x,\hat x)\in R:\mathbf{d}(H(x),\hat H(\hat x))\leq\varepsilon$;	
  \item
  For any $(x,\hat x)\in R$, we have
  \begin{enumerate}
  \item
  $\forall x \rTo^{u} x',\exists \hat x \rTo^{\hat{u}} \hat x':(x',\hat x')\in R$;
  \item
  $\forall \hat x \rTo^{\hat u} \hat x',\exists x \rTo^{u} x':(x',\hat x')\in R$.
  \end{enumerate}
\end{enumerate}
We say that $\Sigma$ is $\varepsilon$-InitSOP simulated by $\hat \Sigma$, denoted by $\Sigma\preceq_I^\varepsilon \hat \Sigma$,
if there exists an $\varepsilon$-InitSOP  simulation relation $R$ from $\Sigma$ to $\hat \Sigma$.
\end{definition}
\end{tcolorbox}
Note that a system $\hat \Sigma$ that simulates $\Sigma$ through the InitSOP simulation relation is often called an opacity-preserving abstraction of $\Sigma$. 
We should mention that, although the above relation appears to be similar to the approximate bisimulation relation proposed in \cite{girard07}, it is still a one-sided relation here because Condition 1 is not symmetric. We refer the interested readers to \cite{Zhang2018OpacitySimilation} to see why one needs the strong Condition 3 in Definition \ref{InitSOP} to show preservation of initial-state opacity in one direction when $\varepsilon=0$. 
Similar notions of approximate simulation relations for preserving current-state and infinite-step opacity are introduced in \cite{Yinapproximate} and omitted here due to lack of space.

The following theorem provides a sufficient condition for verifying $\delta$-approximate initial-state opacity based on related systems as in Definition \ref{InitSOP}.

\begin{tcolorbox}
\begin{theorem}[\textbf{Abstraction-based Opacity Verification}]\label{thm:InitSOP}\upshape
Consider two  metric systems  $\Sigma=(X, X_{0},X_{S},U,f,Y,H)$ and $\hat \Sigma=(\hat X,\hat X_{0},$ $\hat X_{S}, \hat U,\hat f,\hat Y,\hat H)$ with the same output sets $Y= \hat Y$ and metric $\mathbf{d}$ and
let $\varepsilon,\delta\in{R}_0^{+}$.
If  $\Sigma\preceq_I^\varepsilon \hat \Sigma$ and $\varepsilon\leq \frac{\delta}{2}$,
then we have:
\begin{align}
            \hat \Sigma\text{ is ($\delta-2\varepsilon$)-approximate  opaque} \Rightarrow \Sigma \text{ is $\delta$-approximate  opaque}.\nonumber
\end{align}
\end{theorem}
\end{tcolorbox}
Note that the above implication across two related systems holds for all of the three types of approximate opacity.
This result  provides us a sufficient condition for verifying approximate opacity using abstraction-based techniques. 
It is worth remarking that $\delta$ and $\varepsilon$ are parameters specifying two different types of precision.
Parameter $\delta$ is used to specify the measurement precision under which we can guarantee opacity for a single system,
while parameter $\varepsilon$ is used to characterize the ``distance" between two systems in terms of preservation of approximate opacity. 

We illustrate the usefulness of $\varepsilon$-approximate initial-state opacity preserving simulation relation  by the following example. 

\begin{example}\label{exmaple3}
Consider two systems $\Sigma$ and $\hat \Sigma$ as shown in Figure~\ref{fig:4}, where the outputs are specified by the values inside the brackets associated to each state, and secret states are marked in red. 
First note that one can easily verify that the smaller system $\hat \Sigma$ is $\delta$-approximate initial-state opaque with $\delta=0.1$. 
Next, we show that $\Sigma$ is $\varepsilon$-approximate InitSOP simulated by $\hat \Sigma$, as in Definition~\ref{InitSOP}, through the relation
$R=\{(A,J),(B,K),(C,K),(D,K),(E,N),(F,M), (G,$ $M),(I,M)\}$, where $\varepsilon=0.1$.
Condition 1 in Definition~\ref{InitSOP} can be easily checked since : a) for $E\in X_{0}\cap X_{S}$, there exists $N\in  \hat X_{0}\cap \hat X_{S}$ such that $(E,N)\in R$; b) for $J \in \hat X_{0}\setminus \hat X_{S}$, there exists $A\in   X_{0}\setminus X_{S}$ such that $(A,J)\in R$.
Condition 2 is satisfied readily by seeing $\mathbf{d}(H(x),\hat H(\hat x))\leq 0.1$ holds for any $(x,\hat x)\in R$. 
One can also verify that Condition 3 holds as well by checking Conditions 3a) and 3b) for each pair of states in the relation $R$. For instance, consider the state pair $(C,K) \in R$, we have for $C \rTo D$, there exists $K \rTo K$, such that $(D,K) \in R$, and vice versa. Hence, $R$ is an $\varepsilon$-InitSOP simulation relation from $\Sigma$ to $\hat  \Sigma$ as in Definition~\ref{InitSOP}. 
Now, without applying any verification algorithm to $\Sigma$, by leveraging the results in Theorem~\ref{thm:InitSOP}, we can readily conclude that $\Sigma$ is $0.3$-approximate initial-state opaque, where $0.3=\delta+2\varepsilon$.
\begin{figure}[!t]
  \centering
	\includegraphics[width=0.45\textwidth]{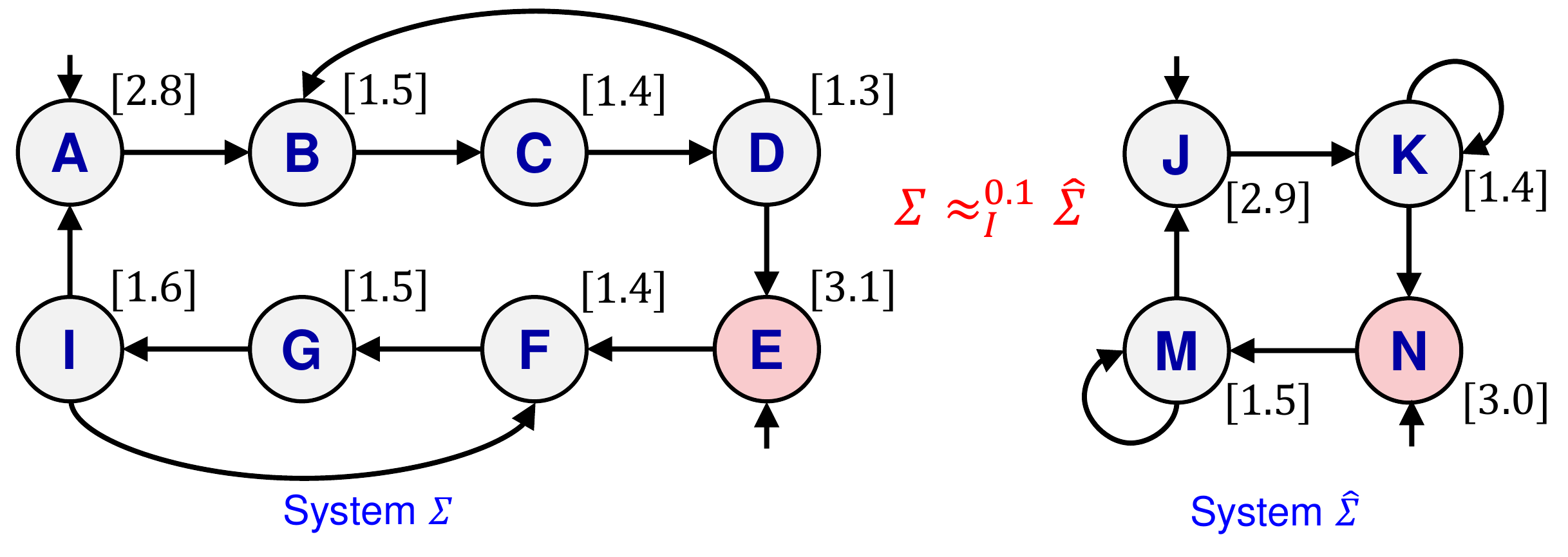}
\caption{Example of $\varepsilon$-approximate initial-state opacity preserving simulation relation.} 
\label{fig:4}
\vspace{-0.3cm}
\end{figure}
\end{example}

Till here, we have introduced  notions of approximate opacity-preserving simulation relations and discussed their properties as in Theorem \ref{thm:InitSOP}. 
As mentioned before, this allows us to verify approximate opacity for infinite systems, e.g., continuous-space control systems, based on their finite abstractions.  
In the following, we present how to construct finite abstractions for a class of dt-CS for the purpose of verifying approximate opacity under the assumption of incremental input-to-state stability ($\delta$-ISS) \cite{angeli}. Formally, a dt-CS $\Sigma$ is called incrementally input-to-state stable ($\delta$-ISS) if there exist a $\mathcal{KL}$ function $\beta$ and $\mathcal{K}_\infty$ function $\gamma$ such that for all $x,x'\in  X$ and for all $\nu,\nu':\mathbb{N}\to  U$, the following inequality holds for any $k\in\N$:
	\begin{align}\label{ISS_enq}
	\Vert \mathbf{x}_{x,\nu}(k)\!-\!\mathbf{x}_{x',\nu'}(k)\Vert\!\leq\!\beta(\Vert x-x'\Vert,k)\!+\!\gamma(\Vert \nu-\nu'\Vert_\infty).
	\end{align} 
Now, consider a concrete control system $\Sigma\!=\!( X, X_0, X_S, U,f, Y,H)$. 
Assume that the output map $H$ satisfies the following general Lipschitz assumption: $\Vert H(x)\!-\!H(x')\Vert\!\leq\! \alpha(\Vert x\!-\!x'\Vert)$,
for all $x,x'\!\in \!X$, where $\alpha\!\in\!\mathcal{K}_{\infty}$.
Consider a tuple $\mathsf{q}=(\eta,\mu)$ of parameters, where $0<\eta\leq\min\left\{\boxspan({X_S}),\boxspan({X}\setminus{X_S})\right\}$ is the state set quantization, and $0<\mu\leq\boxspan({U})$ is the input set quantization parameter. A finite abstraction of $\Sigma$ is defined as 
\begin{equation} \label{symbolicmodel}
\hat \Sigma=(\hat X,\hat X_{0},\hat X_{S}, \hat U,\hat f,\hat Y,\hat H),
\end{equation}
where $\hat X = \hat X_{0}=\left[ X\right]_\eta$, 
$\hat X_{S}=\left[ X_{S}\right]_\eta$, $\hat U =\left[ U\right]_\mu$, $\hat Y=\{H(\hat x)\,\,|\,\,\hat x \in \hat X \}$, where $\hat H(\hat x)=H(\hat x)$, $\forall \hat x\in \hat X$, and
\begin{itemize}
\item $\hat x' \in \hat f(\hat x,\hat u)$ if and only if $\Vert \hat x'-f(\hat x,\hat u)\Vert\leq\eta$.
\end{itemize}

The following result shows that, under some condition over the quantization parameters $\eta$ and $\mu$, $\hat \Sigma$ and $\Sigma$ are related under the approximate InitSOP simulation relation as in Definition \ref{InitSOP}. 
\begin{tcolorbox}
\begin{theorem}[\textbf{Opacity Preserving Finite Abstractions}]\label{theorem1}\upshape
Consider a $\delta$-ISS control system $\Sigma \!=\!( X, X_0, X_S, U,f, Y,H)$. For any desired precision $\varepsilon>0$, let $\hat \Sigma$ be a finite abstraction of $\Sigma$ with a tuple $\mathsf{q}=(\eta,\mu)$ of parameters satisfying
\begin{equation}\label{bisim_cond}
\beta\left(\alpha^{-1}(\varepsilon),1\right)+\gamma(\mu)+\eta\leq\alpha^{-1}(\varepsilon),
\end{equation}
then, we have $\Sigma\preceq_I^\varepsilon \hat \Sigma \preceq_I^\varepsilon \Sigma$. 
\end{theorem}
\end{tcolorbox}

We would like to refer interested readers to \cite[Example. VI.9]{Yinapproximate} for an example that illustrates how to use Theorem~\ref{theorem1} to verify approximate opacity for an infinite system based on its finite abstraction. 

Here, we presented the results mainly tailored to initial-state opacity to illustrate the rough idea of abstraction-based approaches for verifying opacity of continuous-space CPS. Note that similar results on the preservation of approximate current-state and infinite-step opacity through related systems can be found in \cite{Yinapproximate}. 
We would like to refer interested readers to some extensions of the results illustrated above to larger classes of systems including stochastic systems \cite{liu2020notion} and switched systems \cite{liu2021verification,liu2020compositional}.

\subsection{CPS: Deductive Approach via Barrier Certificates}\label{sec:veri_bc}
%
The results discussed in the previous subsection provides a systematic framework to deal with opacity properties for complex CPS. However, this methodology may suffer from scalability issues since it requires discretization of the state and
input sets of the original system. 
As an alternative, there is a growing interest in developing discretization-free approaches for the formal verification of privacy properties based on  notions of \emph{barrier certificates}. In the past decade, barrier certificates have shown to be a promising tool for the analysis of safety problems  \cite{prajna2007framework, ames2016control,ames2017control,wang2017safety} and recently extended to deal with more general temporal logic specifications \cite{jagtap2019formal,lindemann2018control,anand2021formal}. A recent attempt to analyze privacy of CPS using barrier certificates is made in \cite{ahmadi2018privacy}.
A new notion of current-state opacity was considered there
based on the belief space of the intruder.
The privacy verification problem is cast into checking a safety property of the intruder's belief dynamics using barrier certificates. However, this framework is again limited to systems modeled by partially-observable Markov decision processes (POMDPs) with finite state sets.
In this subsection, we revisit a discretization-free approach proposed in \cite{liu2020verification} that is sound in verifying approximate initial-state opacity for discrete-time control systems. 

Consider a dt-CS $\Sigma\!=\!(X, X_0, X_S, U,f, Y, H)$.  
We define the associated augmented system by 
\begin{align} \notag
\Sigma \!\times\! \Sigma\!=\!( X \!\times\!  X,\! X_0 \!\times\!  X_0,\! X_S \!\times\!  X_S, U \!\times\!  U,\!f \!\times\!f, Y \!\times\!  Y,\! H \!\times\! H),
\end{align}
which can be seen as the product of a dt-CS $\Sigma$ and itself. 
For later use, we denote by $(x,\hat x) \!\in\!  X \!\times\!  X$ a pair of states in $\Sigma \!\times \!\Sigma$ and by $(\mathbf{x}_{x_0,\nu},  \mathbf{x}_{\hat x_0,\hat \nu})$ 
the state trajectory of $\Sigma \times \Sigma$ starting from $(x_0, \hat x_0)$ under input run ($\nu,\hat \nu$). We use  $\mathcal{R}\!=\! X \!\times  X$ to denote the augmented state space.
In order to leverage barrier certificates to verify approximate initial-state opacity for a dt-CS $\Sigma$, we further define two sets of interests, i.e., the sets of initial conditions $\mathcal{R}_0$ and unsafe states $\mathcal{R}_u$, as: 
\begin{align}\label{set:initial}
\mathcal{R}_0\!=&\{(x,\hat x) \!\in \!( X_0 \!\cap\!  X_S) \!\times\! ( X_0 \!\setminus \! X_S) \!: \Vert H(x)\!-\!H(\hat x)\Vert \!\leq\! \delta\}, \\ \label{set:unsafe}
\mathcal{R}_u\!=&\{(x,\hat x)\!\in \! X \!\times \! X : \Vert H(x)\!-\!H(\hat x)\Vert \!>\! \delta \},
\end{align}
where $\delta \in {R}_{\geq 0}$ captures the measurement precision of the intruder as introduced in Definition \ref{def:opa-app}.

\begin{figure}
    \centering
 	\includegraphics[scale=0.35]{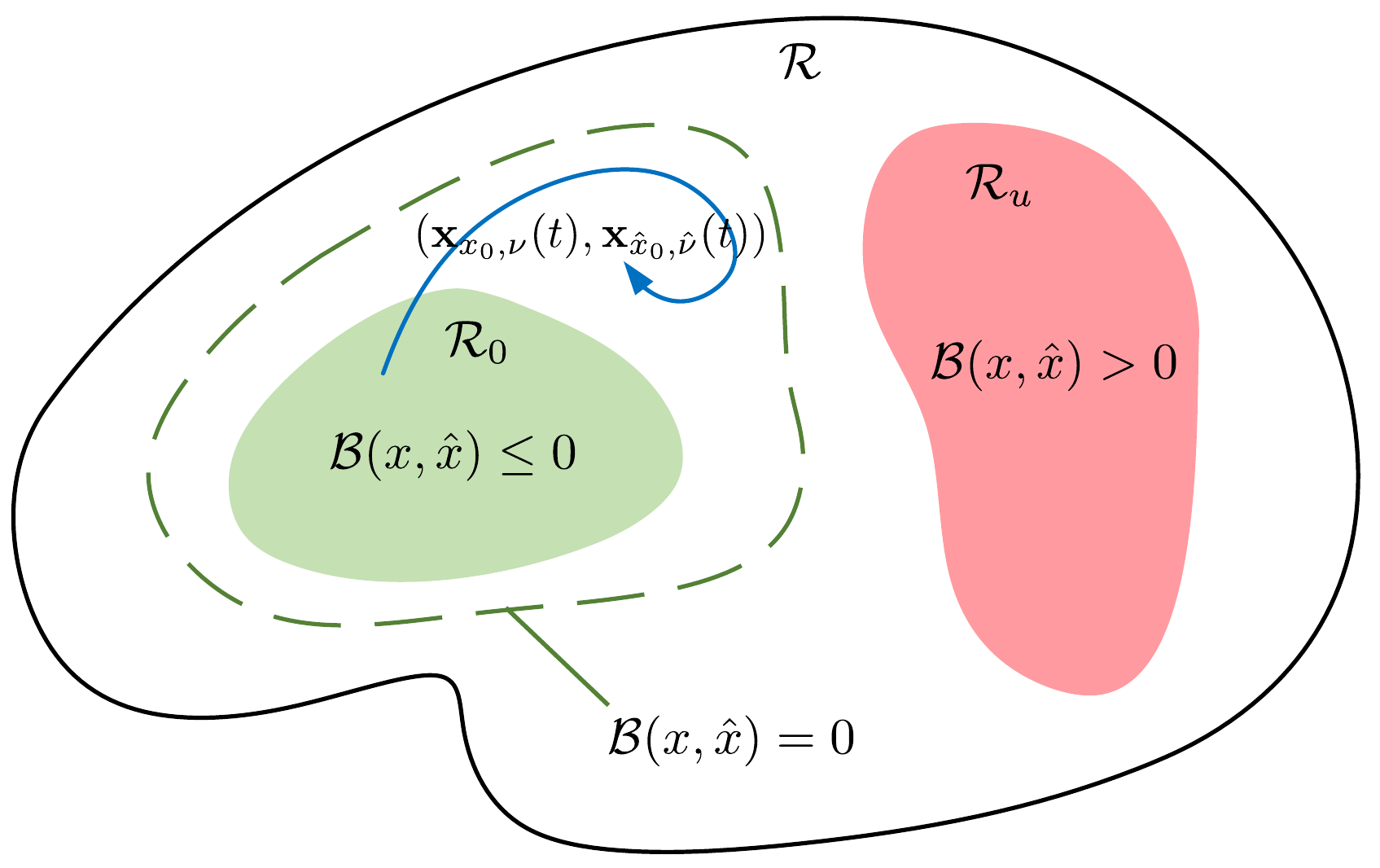}
    \caption{Barrier certificate ensuring safety of the augmented system, which implies opacity of the original system.}
    \label{fig:safety_barrier}
\end{figure}

The following theorem provides a sufficient condition in verifying approximate initial-state opacity of discrete-time control systems via a notion of barrier certificates. 
\begin{tcolorbox}
\begin{theorem}[\textbf{Barrier Certificates for Verifying  Opacity}]\label{BC}\upshape
 	Consider a dt-CS $\Sigma$, the associated augmented system $\Sigma \!\times\! \Sigma$, and sets $\mathcal{R}_0, \mathcal{R}_{u}$ in \eqref{set:initial}-\eqref{set:unsafe}. 
	Suppose there exists a function ${B}: X \times  X \!\rightarrow\! \mathbb{R}$ such that  
	\begin{align*} 
	&\forall (x,\hat x) \in \mathcal{R}_0, \quad \quad \quad \quad \quad  {B}(x,\hat x) \leq 0, \\ 
	&\forall (x,\hat x) \in \mathcal{R}_u, \quad \quad \quad \quad \quad  {B}(x,\hat x) > 0, \\   \notag
	&\forall (x,\hat x) \in \mathcal{R}, \forall u \in {U}, \exists \hat u \in {U}, \\ 
	& \quad \quad \quad \quad  {B}(f(x,u),f(\hat x,\hat u)) - {B}(x,\hat x)\leq 0. 
	\end{align*}
	Then, for any  $(x_0, \hat x_0) \!\in\! \mathcal{R}_0$ and for any input run $\nu$, there exists an input run $\hat \nu$ such that
$(\mathbf{x}_{x_0,\nu}(t), \mathbf{x}_{\hat x_0,\hat \nu}(t)) \cap \mathcal{R}_u = \emptyset$,  $\forall t \in \mathbb{N}$. This implies that $\Sigma$ is $\delta$-approximate initial-state opaque.
\end{theorem}
\end{tcolorbox}
A function ${B}(x,\hat x)$ that satisfies the conditions in Theorem \ref{BC} is called an \emph{augmented control barrier certificate} for $\Sigma \!\times \! \Sigma$. 
This result shows that the existence of such barrier certificates ensures a safety property for $\Sigma \!\times\! \Sigma$, which further implies opacity property of $\Sigma$. 
The interpretation of Theorem \ref{BC} is depicted in Figure.~\ref{fig:safety_barrier}. 
It is worth noting that, failing to find such a barrier certificate does not necessarily imply that the system is not opaque. In this situation, a natural question is whether or not we can use similar barrier-certificates based approaches to show the lack of opacity. 
This problem is addressed in \cite{liu2020verification} and briefly presented next.

\begin{figure}
    \centering
 	\includegraphics[scale=0.35]{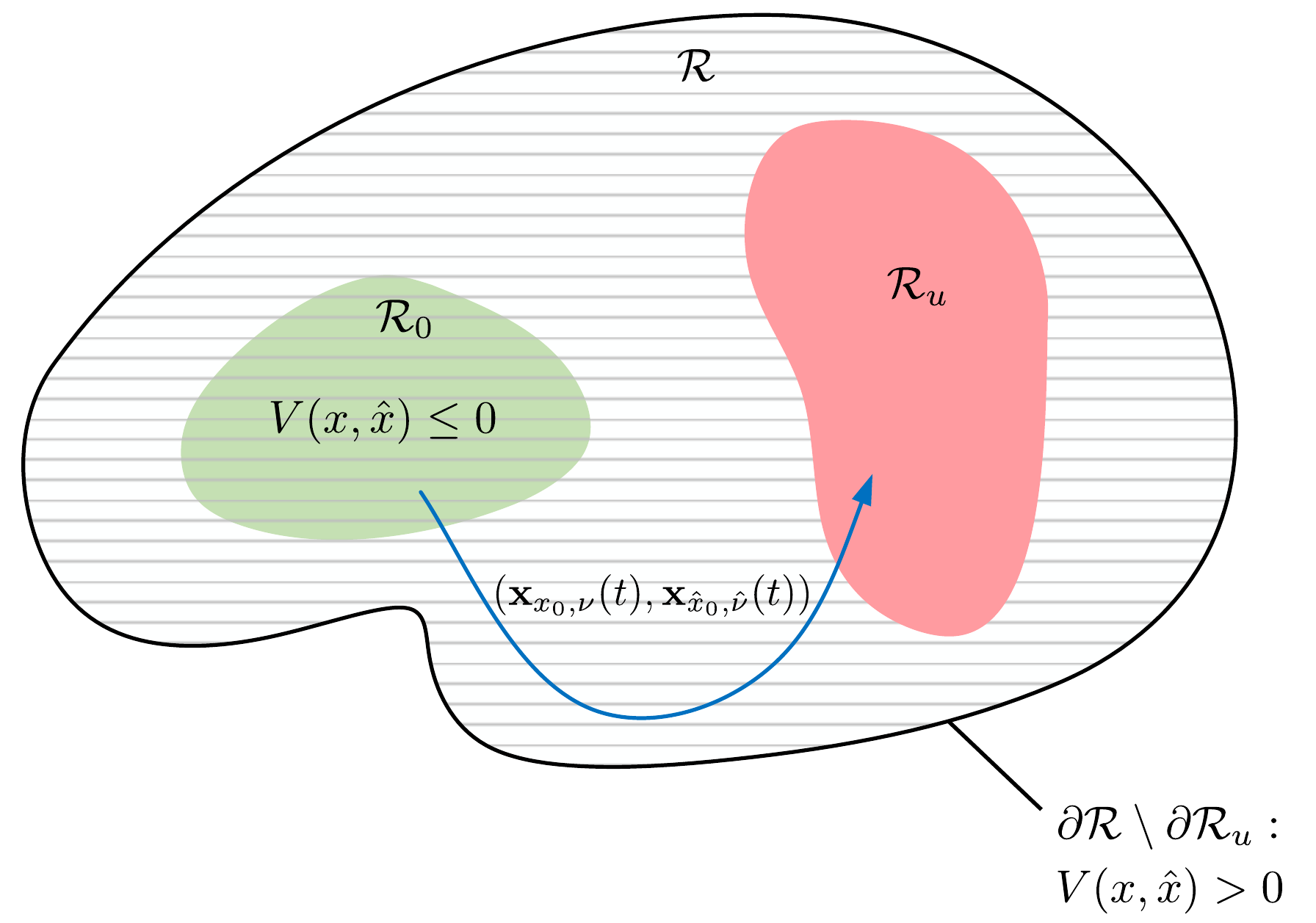}
    \caption{Barrier certificate ensuring reachability of the augmented system, which implies lack of opacity of the original system.}
    \label{fig:reachability_barrier}
\end{figure}

\begin{tcolorbox}
\begin{theorem}[\textbf{Barrier Certificates for Verifying Lack of Opacity}]\label{BC1}\upshape
	Consider a dt-CS $\Sigma$, the associated augmented system $\Sigma \times \Sigma$, and sets $\mathcal{R}_0, \mathcal{R}_{u}$ given in \eqref{set:initial}-\eqref{set:unsafe}. 
Suppose $ X \subset \mathbb{R}^n$ is a bounded set and there exists a continuous function $V: X \times  X \rightarrow \mathbb{R}$ such that 
	\begin{align*} 
	&\forall (x,\hat x) \in \mathcal{R}_0, \quad \quad \quad \quad \quad  V(x,\hat x) \leq 0, \\ 
	&\forall(x,\hat x) \in \partial \mathcal{R} \setminus \partial  \mathcal{R}_u, \quad \quad V(x,\hat x) > 0, \\  \notag
	&\forall(x,\hat x) \in \overline{(\mathcal{R} \setminus \mathcal{R}_u)},  \exists u \in {U}, \forall \hat u \in {U},\\ 
	& \quad \quad \quad \quad  V(f(x,u),f(\hat x,\hat u))-V(x,\hat x) < 0.
	\end{align*} 
Then, for any $(x_0, \hat x_0) \in \mathcal{R}_0$, there exists an input run $\nu$ such that $(\mathbf{x}_{x_0,\nu}(T), \mathbf{x}_{\hat x_0,\hat \nu}(T)) \in \mathcal{R}_u$ for any $\hat \nu$, for some $T \geq 0$, and $(\mathbf{x}_{x_0,\nu}(t), \mathbf{x}_{\hat x_0,\hat \nu}(t)) \in \mathcal{R}$, for all $t \in [0,T]$.
This implies that system $\Sigma$ is not $\delta$-approximate initial-state opaque. 
\end{theorem}
\end{tcolorbox}
In particular, the previous theorem provides a sufficient condition to verify the lack of approximate initial-state opacity by constructing another type of augmented control barrier certificates ensuring a reachability property for $\Sigma \!\times\! \Sigma$. 
The interpretation is illustrated in Figure.~\ref{fig:reachability_barrier}.

We should mention that, by defining proper regions of interest, i.e., the sets of initial conditions $\mathcal{R}_0$ and unsafe states $\mathcal{R}_u$ for the barrier certificates, similar results can be derived for the verification of other types of approximate opacity; see, e.g., \cite{kalat2021modular}.


For systems with polynomial transition functions and semi-algebraic  sets (i.e., described by polynomial equalities and inequalities) ${X}_0$, ${X}_S$, and ${X}$, 
an efficient computational method based on sum-of-squares (SOS) programming can be utilized to search for polynomial barrier certificates.
In this way, one can leverage existing computational toolboxes such as \texttt{SOSTOOLS} \cite{SOSTOOLS} together with semidefinite programming solvers such as \texttt{SeDuMi} \cite{sturm1999using} to compute polynomial barrier certificates. We refer interested readers to \cite[Sec. IV]{liu2020verification} for more details on how to translate barrier conditions to SOS constraints.
Note that by formulating the barrier conditions as a satisfiability problem, one can alternatively search for parametric control barrier certificates using an iterative program synthesis framework, called Counter-Example-Guided Inductive Synthesis (\texttt{CEGIS}), with the help of Satisfiability Modulo Theories (\texttt{SMT}) solvers such as \texttt{Z3} \cite{de2008z3} and \texttt{dReal} \cite{gao2013dreal}; see, e.g., \cite{jagtap2019formal} for more details. We also refer interested readers to the recent work \cite{peruffo2020automated}, where machine learning techniques were exploited for the construction of barrier certificates.

\subsection{Ongoing \& Open Problems}
So far, we discussed the basic security verification procedures for general CPS using abstractions and barrier certificates.  In the followings, we further discuss some ongoing research topics and open problems.

\paragraph{\emph{Verification of General Notion of Opacity for CPS}} Existing works for opacity verification of general CPS mainly focus on particular types of opacity such as initial-state opacity or infinite-step one. For finite systems, the general notion of $\alpha$-opacity as defined in Definition~\ref{def:op-general} can be verified using the observer-like structures when the security properties can be realized by $\omega$-automata. However, for general CPS with infinite states, how to verify the general notion of $\alpha$-opacity   still needs developments.  In particular, for the abstraction-based approach, one needs to identify suitable relation that preserves $\alpha$-opacity.  For the barrier-based approach, appropriate conditions for barrier certificates of $\alpha$-opacity also need to be identified. 

\paragraph{\emph{Quantitative Verification of Opacity}}
The opacity verification problem discussed in this section is \emph{binary} in the sense that  the system is either opaque or not. In some cases, when the verification result is negative, one may be further interested in how insecure the system is. This motivates the research of \emph{quantifying} the level of information leakage. For finite systems, one popular approach is to consider systems modeled by probabilistic finite-state automata, Markov chains or Markov decision processes. Then one can quantify opacity in terms of probability  \cite{saboori2014current,berard2015probabilistic,berard2015quantifying,keroglou2018probabilistic,yin2019infinite,lefebvre2021exposure}. 
For example, one may require that the intruder can never know that the system is currently at a secret-state with more than $\epsilon$ probability, or the system has less than $\epsilon$ probability to reveal its secret. 
However, all existing works on quantifying opacity consider finite systems, although their belief spaces may be infinite. How to leverage opacity quantification techniques for general CPS, using  either abstraction-based approaches  or barrier certificates, still need to be developed. The recent result in \cite{liu2020notion} has made some initial steps towards this objective using the abstraction-based technique. 

\paragraph{\emph{Opacity Verification for Larger Classes of CPS}} 
The aforementioned abstraction-based approaches for opacity verification of general CPS crucially depends on incremental ISS assumption. However, this assumption is rather restrictive for many practical systems. How to relax the stability assumption so that the verification techniques can be applied to more general classes of CPS is an interesting and important future direction.  
 
Also, in the problem formulation of opacity, the attacker is assumed to be able to access partial information-flow of the plant. However, for networked control systems, the information transmission between controllers and plants in the feedback loops  may also be released to the intruder. There are some very recent works on the verification of opacity for  networked control systems using finite-state models; see, e.g., \cite{yin2018verification,yang2021opacity,zhang2021networked,lin2020information,yangjk2021opacity}. However, 
existing works on formal verification of networked control system mainly focus on the mission requirements \cite{zamani2018symbolic,hashimoto2019symbolic,pola2019control,borri2019design} and to the best of our knowledge, there is no result on formal verification of opacity for general  networked CPS.
 
\section{Secure-by-Construction Controller Synthesis}

In the previous section, we investigated the security verification problem for open-loop systems.   However, the original system $\Sigma$  may not be opaque. 
Therefore, it is desired to \emph{enforce} opacity for the system via the feedback control mechanism. 
In the realm of control theory, one of the most popular approaches for
enforcing certain property of the system is through a feedback controller.

A {\it supervisor} or a controller  for $\Sigma$ is a function 
$C:\texttt{Path}(\Sigma)\to 2^U$ that determines a set of possible control inputs  based on the executed state sequences. 
We denote by $\Sigma_C$ the closed-loop system under control. 
Specifically, a state run $x_0\rTo^{u_1}x_1\rTo^{u_2}\cdots\rTo^{u_{n-1}}x_{n-1}\rTo^{u_{n}}x_n$ is feasible in the closed-loop system if it is a run in the open-loop system $\Sigma$ and   $u_i\in C(x_0x_1\cdots x_{i-1})$ for any $i\geq 1$. 
Similarly, we denote by $\texttt{Path}^{(\omega)}(\Sigma_C)$ and $\texttt{Trace}^{(\omega)}(\Sigma_C)$ the set of paths and the set of traces of the controlled system  $\Sigma_C$, respectively.
 
The goal of the control synthesis problem is to synthesize a feedback controller
$C$ such that the closed-loop system $\Sigma_C$ satisfies both the mission requirement, e.g.,  an LTL formula $\varphi$,
and/or, the security requirement, e.g.,  opacity.
Specifically, we investigate the following control synthesis problem. 

 \begin{tcolorbox}
 \begin{problem}[\textbf{Secure-by-construction Controller Synthesis}]\label{prob:synthesis}\upshape
 Given a mission requirement (as an LTL formula) $\varphi$ and a security property $\alpha$, the secure-by-construction controller synthesis problem is to design a supervisor $C$ such that $\Sigma_C \models(\varphi, \alpha)$.
 \end{problem}
 \end{tcolorbox}

The foundations for the {\it correct-by-construction} approach were laid by
Church in~\cite{Church1963} where he stated his famous {\it synthesis problem}:
{\it given a requirement which a circuit is to satisfy, find a circuit that
  satisfies the given requirement (or alternatively, to determine that there
  is no such circuit)}.  
The landmark paper by B\"uchi and Landweber~\cite{BL69} gave the first solution of
Church's synthesis problem for specification given in Monadic second-order logic. 
Pnueli and Rosner~\cite{PR89} studied the synthesis problem for specifications
given as LTL~\cite{katoen08} and showed the problem
to be complete with $\textsc{2Exptime}$ complexity. Ramadge and Wonham~\cite{ramadge} studied the synthesis problem---as a mechanism
for supervisory controller synthesis of discrete event systems---for simple {\it
  safety specifications} and gave an efficient linear-time algorithm for
computing maximally permissive controller for this fragment. 
The relation between reactive synthesis and supervisory control has been thoroughly
discussed in a serious of recent works; see, e.g., \cite{ehlers2017supervisory,ramezani2019comparative,partovi2019reactive,schmuck2020relation,majumdar2022supervisory,sakakibara2022finite}. 
The goal of this thrust is to study decidability and complexity of the synthesis
problems for LTL specification (and their efficiently solvable sub-classes) with
security requirements and propose efficient algorithms to solve synthesis
problems.

\subsection{Finite Systems}

In opacity enforcement using supervisory control, the objective is to synthesize a supervisor $C$ that avoids executing those ``secret-revealing" paths and at the same time, satisfies the desired mission requirement described as an LTL formula.   
Note that in Problem~\ref{prob:synthesis}, the meaning of mission satisfaction, i.e.,  $\Sigma_C\models\varphi$, is relatively clear. However, there  may have different interpretations for security for the closed-loop system, i.e., $\Sigma_C\models \alpha$.  
In particular, the synthesis problem can be categorized as policy-aware synthesis and policy-unaware synthesis. 
Here, we still use initial-state opacity as the concrete security property to illustrate the differences. 

\paragraph{\emph{Basic Opacity-Enforcing Controller Synthesis Problem}}  
The most basic setting for opacity enforcing control is to assume that the intruder is \emph{not aware} of the presence of the controller $C$.  
In this setting, we say controller $C$ enforces initial-state opacity for system $\Sigma$ if  
for any  path $\mathbf{x}=x_0x_1\cdots x_n\in \texttt{Path}(\Sigma_C)$, where  $x_0\in X_S$, 
there exists a path $\mathbf{x}'=x_0'x_1'\cdots x_n'\in \texttt{Path}(\Sigma)$, where  $x_0'\notin  X_S$, such that $H(\mathbf{x})=H(\mathbf{x}')$.  
Note that, here, the first secret path $\mathbf{x}$ belongs to the closed-loop system $\Sigma_C$ since we consider those secret paths that can actually happen. 
However, the second non-secret path $\mathbf{x}'$ belongs to the open-loop system $\Sigma$ as we assume that the intruder is unaware of control $C$. 

The basic idea for solving the basic synthesis problem is to construct the corresponding (initial, current or delayed) state-estimator $Obs(\Sigma)$  based  on the open-loop system $\Sigma$.  Then we compose the system $\Sigma$, the state-estimator $Obs(\Sigma)$ and the deterministic Rabin automata  for $\varphi$ to obtain a new system $\Sigma'$.  Then controller $C$ can be synthesized by solving a Rabin game   over $\Sigma'$ for the Rabin acceptance condition \cite{gradel2002automata} and at the same time avoiding reaching those secret-revealing estimator states in $Obs(\Sigma)$. Complete solution for this problem can be found in \cite{takai2008formula,tong2018current,xie2021secure,ma2021verification}; some of them do not consider the LTL mission requirement, which can be addressed  easily by combining with the standard LTL synthesis procedures. 

\paragraph{\emph{Policy-Awareness and Imperfect Information}}
The above basic synthesis problem is  based on the assumptions that (i) the controller has full state information; and (ii) the intruder is unaware of the implementation of the controller. In particular, the latter assumption is reflected by the fact that we choose  non-secret path $\mathbf{x}'$ from the original open-loop system $\texttt{Path}(\Sigma)$ rather than the closed-loop one $\texttt{Path}(\Sigma_C)$.  
However, in practice,  the control policy may become a public information, which is also available to the intruder.  
Then the intruder may further use the knowledge of the controller to improve its state estimator, e.g., it can exclude some paths that have already been disabled by the controller during the state estimation process. 
In order to ensure opacity for this general case, one needs to further investigate how control affects estimation in the synthesis phase. That is, the state estimate of the intruder cannot be constructed solely based on the original open-loop system but should also based on the synthesized control policy. Interested readers are referred to \cite{dubreil2010supervisory,saboori2011opacity,yin2016uniform,xie2021opacity} for the complete solution to this general case for finite systems.  
 
Another practical design consideration is the  imperfect information of the controller.  
In practice, the controller also may not be able to access the full state information of the system. 
Instead, the controller may have its own observation specified by a new output mapping $H_C: X \to O$ and a controller with imperfect information is a function of the form $C:O^*\to 2^U$, which determines the control input based on its own observation. 
Systematic procedures for synthesizing controllers under imperfect information can be found in \cite{arnold,raskin2007algorithms,thistle2009effective,yin2016tac}. 
In the context of opacity-enforcing synthesis, the main difficulty here is that the information of the intruder and the information of the controller may be \emph{incomparable}, i.e., the equivalent classes induced by mappings $H$ and $H_C$ are incomparable. Interested readers are referred to \cite{dubreil2008opacity,dubreil2010supervisory} for more discussions on this issue.

\paragraph{\emph{Opacity-Preserving Path Planning}} 
The complexity of the basic opacity-enforcing controller synthesis problem is exponential in the size of $\Sigma$ due to the subset construction used in the state estimators 
and double-exponential in the length of the LTL formula $\varphi$ due to the construction of the deterministic Rabin automaton.  
Note that one has to use deterministic $\omega$-automata to realize the LTL formulae because the plant under control is non-deterministic in general. 
However, when system $\Sigma$ is deterministic, the basic  synthesis problem becomes a \emph{planning problem} for which the computational complexity can be significantly improved. 
In particular, when system $\Sigma$ is deterministic, a deterministic controller is also referred to as a \emph{plan} because the trajectory of the system can be completely determined without any uncertainty. Therefore, for planning problem, one just needs to find an infinite path satisfying both the mission and the security requirements.  
The results in \cite{hadjicostis2018trajectory} investigate the problem of planning a trajectory towards a target state  under current-state opacity constraints. 
The results in \cite{yang2020secure} consider the security-aware path planning problem together with LTL mission requirements. The idea is to construct the so-called twin-system, whose size is polynomial with respect to the size of $\Sigma$, to capture the security requirements without building the exponentially large state estimator. Furthermore, since system $\Sigma$ is already deterministic, one can further use non-deterministic B\"{u}chi automata, whose size is single-exponential in the length of formula $\varphi$, to capture the LTL specification.  In this case, the complexity of the opacity synthesis can be reduced to polynomial in the size of the system and to single-exponential in the length of $\varphi$. 

\paragraph{\emph{Other Opacity Enforcement Mechanisms}} 
In the above paragraphs, we discussed the enforcement of opacity using feedback controllers. 
In some applications, however, one cannot change the actual behavior of the system directly. 
Therefore, many different alternative enforcement mechanisms have also been developed by changing the information-flow available to the intruder to ensure security of the systems.  
For example, in \cite{wu2014synthesis,ji2019enforcing,wubo2018synthesis}, \emph{insertion functions} were used to ``confuse" the intruder by adding factitious symbols to the output sequences. 
Insertion functions have been further generalized to \emph{edit functions} that allow not only event insertions, but also event erasures and replacements \cite{wu2018synthesis,ji2019opacity}. 
Another widely used approach is to synthesize \emph{dynamic masks} \cite{cassez2012synthesis,zhang2015maximum,behinaein2019optimal,yin2019general,yin2020synthesis} that determine  which information  to be released to the outside world under the security constraints. 
Other approaches for enforcing opacity include using run-time techniques \cite{falcone2015enforcement} and event shuffles \cite{barcelos2021enforcing}.

\subsection{Secure-by-Construction Controller Synthesis for CPS}
The above discussed controller synthesis techniques are developed for finite systems.  
Those techniques, in general, are not appropriate for CPS with continuous-space dynamics such as systems in the form of equation~\eqref{def:sys1}. Unfortunately, there are only very few recent works on the enforcement of opacity for CPS, which are discussed as follows.

\paragraph{\emph{Abstraction-Based Synthesis}} 
\begin{figure}
  \centering
  \includegraphics[width=0.45\textwidth]{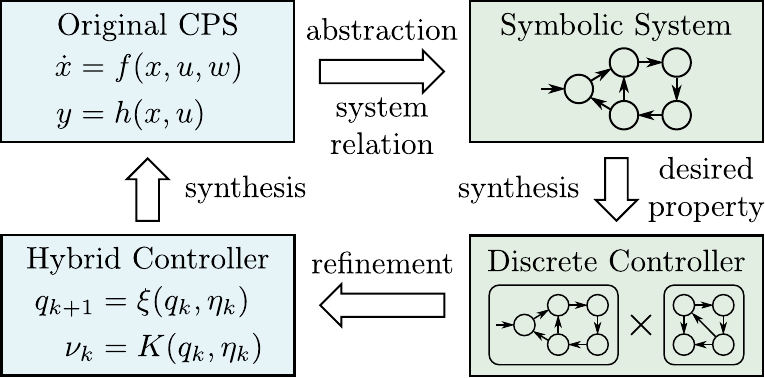}
  \caption{Pipeline of standard discretization-based synthesis technique.}\label{fig:abs_synthesis_pipeline}
\end{figure}
The basic pipeline of abstraction-based or discretization-based controller synthesis is shown in Figure~\ref{fig:abs_synthesis_pipeline}.
Similar to the abstraction-based verification,  
in abstraction-based synthesis, one needs to first build the finite abstraction of the concrete CPS, and then  synthesize a controller based on the finite abstraction, and finally, refine the synthesized discrete controller back as a hybrid controller to the original CPS. 
Then the key question is still to find appropriate relations between  concrete systems and their finite abstractions such that properties of interest can be preserved under controller refinement.  

It is well-known that the (bi)simulation relation is not suitable for the purpose of controller synthesis because it does not take the effect of control non-determinism into account \cite{pola2009symbolic}. To address this issue, one needs to extend the (approximate)  (bi)simulation relations to the (approximate) \emph{alternating} (bi)simulation relations \cite{alur1998alternating,tab09}. However, although the standard alternating simulation relations preserve the LTL mission requirements, they do not preserve security requirements. In \cite{hou2019abstraction},  two notions of \emph{opacity-preserving} alternating simulation relations are proposed, one for initial-state opacity and one for infinite-step opacity.
Based on these notions,  one can synthesize opacity-enforcing controllers  directly by applying existing synthesis algorithms to the finite abstractions that \emph{opacity-preserving alternatively simulate} the concrete systems. 
In \cite{mizoguchi2021abstraction}, the authors propose a two-stage approach for enforcing opacity for CPS. First, a controller ensuring the LTL mission requirement is synthesized based on the standard alternating simulation relations  without considering  opacity. Then those actions violating opacity are eliminated by a symbolic control barrier function such that security requirement is fulfilled.  
  
\paragraph{\emph{Abstraction-Free Synthesis}} 
In the context of discretization-free approaches, to the best our knowledge, only the results in \cite{an2019opacity} investigated the opacity enforcement problem for  restricted classes of CPS and security notions. 
Specifically, they considered CPS modeled by linear time-invariant (LTI) systems and the security requirement is to make sure that the  interference attenuation capacity of the system is opaque. Then the opacity enforcement problem is formulated as an $\mathcal{L}_2$-gain optimization problem for LTI systems. An approximated-based adaptive dynamic-programming (ADP) algorithm was proposed to design an opacity-enforcing controller. 
 
\subsection{Ongoing  \& Open Problems}
In the following, we mention some ongoing research directions and open problems regarding secure-by-construction controller synthesis. Compared with   security-aware verification, secure-by-construction synthesis is less tackled in the literature.

\paragraph{\emph{Synthesis for Finite Systems}}
The opacity-enforcing control problem for finite systems has already been studied for about fifteen  years. However, all existing solutions are either based on the assumption that the knowledge of the supervisor and the intruder are comparable \cite{dubreil2010supervisory,saboori2011opacity,yin2016uniform}, or based on the assumption that the intruder is unaware of the presence of the supervisor \cite{takai2008formula,tong2018current}. The general opacity-enforcing   control problem without any assumption, to the best of our knowledge, is still open even for finite systems. Also, for networked control systems  with both control and observation channel information leakages, 
how to synthesize opacity-enforcing controllers  is still an open problem; so far, only the verification problem is solved for finite systems \cite{yin2018verification,yang2021opacity}. Furthermore, existing works on opacity-enforcing control mainly consider centralized control architectures. In general, the plant may be controlled by a set of \emph{local controllers} with or without communications, which leads to the  distributed \cite{barrett2000decentralized,kalyon2014symbolic} or the decentralized control architectures \cite{yoo2002general,pola2018decentralized}.   How to synthesize opacity-enforcing controllers under those general information structures is still an open problem. 

The high complexity or even undecidability are the major obstacles towards automated controller synthesis of opacity. To overcome this challenge, a potential future direction is to develop {\it bounded-synthesis}~\cite{Schewe07}  that reduces the search for a bounded size
implementation satisfying the synthesis objective to a SAT problem. The key
advantage of the bounded synthesis over traditional synthesis is that it constructs minimal size supervisors.
Therefore, it is a promising direction to extend the bounded synthesis approach to solve controller synthesis problem for generalized
language-based opacity by implicitly encoding the self-composition of the abstract model.  
Another premising direction is to investigate security-aware synthesis for
well-behaved sub-classes of LTL such as Generalized reactivity(1) (GR(1))~\cite{PPS06,bloem2012synthesis}. 
These are sub-classes of the form
\[
(\always \eventually
p_1 \wedge \ldots \wedge \always \eventually p_m) \implies (\eventually \always
q_1 \wedge \ldots \wedge \eventually \always q_n),
\]where $p_i,q_j$, $i\in\{1,\ldots,m\}$, $j\in\{1,\ldots,n\}$, are some predicates.
For GR(1) formulae, Piterman et al.~\cite{PPS06} showed that
synthesis can be performed in (singly) exponential time. Moreover, authors
argued that GR(1) formulas are sufficiently expressive to provide complete
specifications of many designs.
It is promising to develop an analogous result for security-aware controller-synthesis
w.r.t generalized language-based opacity properties. 
 
\paragraph{\emph{Abstraction-Based Synthesis for CPS}} 
The notions of opacity-preserving alternating  simulation relations (ASR) proposed in \cite{hou2019abstraction}
made the first step towards abstraction-based opacity synthesis for CPS. However, it has many limitations that need to be addressed in the future.   
First, the results in \cite{hou2019abstraction} are developed for particular types of state-based opacity. Similar to the verification problem, we also need to extend the results, particularly the underlying simulation relations, to the general case of  $\alpha$-opacity. 
Second, the opacity-preserving ASR belongs to the category of exact simulation. This condition, in general, is too strong for general CPS with continuous state-space. It is likely that there does not exist a finite symbolic model simulating the concrete system exactly. One possible direction to address this issue is to enforce approximate opacity rather than the exact version. To this end, one needs to consider the approximate ASR   \cite{pola2009symbolic,zamani2012symbolic} rather than the exact ASR.  
Third, existing results  only support state-feedback controllers, i.e., the controller knows the current-state of the system precisely. As we discussed, an opacity-enforcing controller is observation-based in general.  
To address this issue, a possible solution is to use the output-feedback refinement relation (OFRR) \cite{reissig2017feedback,khaled2020output} instead of the ASR. How to suitably generalize the OFRR to preserve opacity is still an open problem. 
Finally, although opacity-preserving relations have been identified, there is no abstraction algorithm available so far for building finite abstractions based on the concrete systems with continuous-space dynamics that satisfy those relations. 
When the concrete system is $\delta$-ISS, the abstraction can be done analogous to the case of verification. The major open problem is how to build opacity-preserving finite  abstractions for the purpose of control without the stability assumption. 

\paragraph{\emph{Abstraction-Free Synthesis for CPS}}
As we have already mentioned, there are very few results for abstraction-free opacity synthesis. One important direction is to extend the barrier-certificates techniques for opacity verification  to   opacity synthesis. To this end, one may borrow the idea of control barrier functions \cite{ames2017control,santoyo2021barrier} that generalize the idea of barrier certificates to control systems by explicitly taking the effect of control choices into account.  
Another widely used abstraction-free technique for formal synthesis  is the sampling-based approaches \cite{vasile2013sampling,kantaros2019sampling,luo2021abstraction}.
In this approach, one can use the concrete models of CPS to randomly generate sample paths until a satisfiable path is found. This avoids discretizing the state-space explicitly and under certain conditions, can provide probabilistic complete solutions. 
However, existing sampling-based planning techniques can only handle LTL mission requirements. How to incorporate the security requirements into the sampling-based process needs further developments. 
  














\section{Compositional Reasoning for Scalability}
In the previous sections, we presented various discretization-based and discretization-free approaches in verifying or enforcing opacity and mission requirements for CPS. 
Though promising, when confronted with large-scale interconnected systems, the aforementioned results in general suffer from the so-called the \emph{curse of dimensionality}.
This prevents current techniques from providing automated verification or synthesis for large-scale interconnected
CPS. This is not just a theoretical concern, many safety-critical applications, such as traffic network, automated
highway driving, building management systems, power networks, air traffic management, uninhabited
aerial vehicles, and so on, consist of many subsystems interacting with each other.
One way to address the inherent difficulty in analyzing or controlling complex, large-scale, interconnected systems,
is to apply a ``divide and conquer” strategy, namely, compositional approaches. 

In the past decades, many potential compositionality results have been proposed to tackle the acute computational bottlenecks in the analysis of safety properties for large-scale continuous-space systems \cite{tazaki1,pola16,kim2,kim2018constructing,boskos2015decentralized,rungger2016compositional,swikir2018dissipativity,swikir2019compositional,lavaei2020compositional,liu2021symbolic}.
However, in the context of analyzing security properties, compositional approaches have been explored only recently for modular verification and synthesis of DES in \cite{saboori2010reduced,noori2018compositional,mohajerani2019transforming,tong2019current,yang2021current,zinck2020enforcing}
and for continuous-space systems in \cite{liu2021compositional,liu2021verification,kalat2021modular}.

\subsection{Modular Approaches for Finite Systems}\label{subsec:modular}

Formally, an interconnected large-scale system $\Sigma$ consists of a set of subsystems or local modules $\{\Sigma_1,\dots,\Sigma_n\}$ whose connectivities are specified by an interconnection mapping $\mathcal{I}$. 
In the context of finite systems or discrete-event systems, the interconnection mapping is usually simplified as the synchronization product $\otimes$ over shared events. That is, the monolithic system is  $\Sigma=\Sigma_1\otimes \cdots \otimes\Sigma_n$.  

\paragraph{\emph{General Complexity Results}}
In the context of opacity verification, 
it was first shown by \cite{yin2017verification2} that verifying opacity for modular systems in the form of $\otimes_{i=1}^n\Sigma_i$ is PSPACE-hard. 
This complexity result was then further improved by \cite{masopust2019complexity} to EXPSPACE-complete, which says that the time-complexity for verifying opacity for modular systems  grows \emph{double-exponentially}  fast when the number of subsystems increases. Therefore, verifying opacity directly by computing the entire monolithic model is computationally intractable in general.  Since the opacity synthesis problem is even more difficult than the verification one, its complexity is at least EXPSPACE-hard. 

\paragraph{\emph{Modular Verification}}
The first modular approach for opacity verification was provided in \cite{saboori2010reduced}. Specifically, it identified a structural sufficient condition such that events shared by each pair of subsystems are pairwise observable. With this structural condition, the verification of opacity for system $\otimes_{i=1}^n\Sigma_i$ can be divided as $n$ local verification problems for subsystems $\Sigma_i$, 
which reduces the double-exponential complexity $2^{|X|^n}$ to single-exponential complexity $n2^{|X|}$, where $|X|=\max_{i=1,\dots, n}|X_i|$. 
More recently, the results in \cite{tong2019current,yang2021current} follow the similar line of reasoning by identifying sufficient conditions under which current-state opacity can be verified efficiently using modular approach without building the monolithic system. In \cite{noori2018compositional}, a compositional abstraction technique was developed based on a notion of visible bisimulation relation. This approach was applied to opacity verification of modular systems by incrementally building the monolithic system while avoiding irrelevant components for the purpose of verification. Finally, the results in  \cite{mohajerani2019transforming} investigated how to transform the opacity verification problem for modular systems to a non-blockingness verification problem, for which mature modular verification algorithms have been developed already \cite{mohajerani2016framework}. 

\paragraph{\emph{Modular Synthesis}}
Similar to the verification problem, the existing opacity enforcing synthesis algorithms also need the monolithic model of the system. 
The results in~\cite{zinck2020enforcing} investigated the opacity enforcing controller synthesis for modular systems under the assumption that the attacker can observe the
interface between each local module. Under this assumption, opacity-enforcing controllers $C_i$ can be synthesized for subsystems $\Sigma_i$ individually and  the overall control system $\otimes_{i=1}^n \Sigma_{i,C_i}$ is guaranteed to be opaque.  In \cite{mohajerani2020compositional}, a
compositional and abstraction-based approach is proposed for synthesis of edit functions for opacity enforcement. The idea is similar to \cite{mohajerani2019transforming} and is based on transforming the opacity synthesis problem to an existing supervisor synthesis problem for modular system without security considerations \cite{mohajerani2014framework}. Note that, different from a supervisory controller, an edit function  can only change  the observation of the system and not the actual behavior of the system.

\subsection{Modular Verification for Large-scale CPS: An Abstraction-based Approach}\label{sec:compo_abs}


\begin{figure}
    \centering
 	\includegraphics[scale=1]{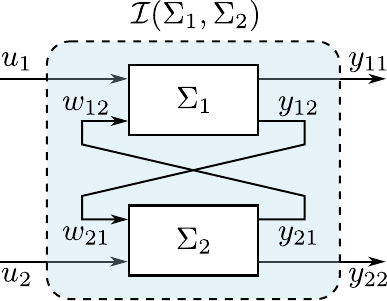}
    \caption{Feedback composition of two subsystems.}
    \label{fig:smallgain}
     \vspace{-0.3cm}
\end{figure}

As we have discussed in Section \ref{Sec:veri_abs}, opacity-preserving finite abstractions and simulation relations serve as a bridge between continuous-space CPS and existing  verification or synthesis algorithms for opacity developed in DES community. Although they are shown to be a useful tool in some recent results \cite{Yinapproximate}, a non-negligible challenge lies in scaling the approach for large-scale systems. Typically, existing techniques reported in Section \ref{Sec:veri_abs} take a monolithic view of systems where abstraction, verification, and synthesis are performed
for the entire system. This monolithic view interacts poorly with the construction of finite abstractions where
the complexity of the construction grows exponentially in the number of state variables in the model. 
Different compositional approaches have been proposed in the literature to overcome this challenge in dealing with large-scale CPS. 
The two most commonly used schemes are based on: 1) assume-guarantee contracts \cite{kim2017small,saoud2021assume,sharf2021assume} which are originally introduced in the computer science literature and 2)  the input-output properties of the system, including those expressed as small-gain \cite{rungger2016compositional,kim2017small,pola16} or dissipativity properties \cite{Zamani18,swikir2018dissipativity} which are originally introduced in the control theory literature. 
Here, the overall large-scale systems are usually seen as  interconnections of smaller (reasonably sized) components, i.e., subsystems. Subsequently, the analysis and the design of the overall system is reduced to those of the subsystems.

In the following, we denote a discrete-time control subsystem by a tuple $\Sigma_i = ( X_i, X_{0_i}, X_{S_i}, U_i, W_i,f_i, Y_i, H_i)$. The formal definition of a control subsystem is similar to the one in \eqref{def:sys1} but with two sets of inputs. In particular, $w \in W$ are termed as ``internal" inputs which are used to describe the interaction between subsystems, and $u \in U$ are called ``external" inputs served as interfaces for controllers.
An interconnected control system composed of $N \!\in\! \N_{\geq 1}$ subsystems is iteself a discrete-time control system as in \eqref{def:sys1}, denoted by $\mathcal{I}(\Sigma_1,\ldots,\Sigma_N)$, subject to certain interconnection constraints. An example of an interconnected system composed of two subsystems is depicted in Figure~\ref{fig:smallgain}.
Now, we briefly discuss a recent result developed in \cite{liu2021compositional} on the compositional construction of opacity-preserving finite abstractions for large-scale CPS.
In order to illustrate the main idea, let us consider the interconnected system depicted in Figure~\ref{fig:smallgain}, which is a feedback composition of two subsystems $\Sigma_1$ and $\Sigma_2$. Suppose each subsystem is denoted by $\Sigma_i = ( X_i, X_{0_i}, X_{S_i},  \varnothing, W_i,f_i,X_i, \mathrm{id})$ and for simplicity described as a  discrete-time linear system:
\begin{align}\label{sys:linear}
	\Sigma_i:\left\{
	\begin{array}[\relax]{rl}
		{x}_i^+ = & \!\!a_i{x}_i+ b_i{x}_j,\\
		{y}_i = &\!\!{x}_i,
	\end{array}\right.
\end{align}
where $\vert a_i\vert<1$. 
Let us define so-called gain functions $\gamma_{i} = \vert b_i/(1-a_i)\vert$ for each $\Sigma_i$. 
The main compositionality result of \cite{liu2021compositional} for this particular setting is summarized as follows.
\begin{tcolorbox}
\begin{theorem}[\textbf{Compositional Construction of Opacity Preserving Finite Abstractions}] \label{Main}\upshape
Consider the interconnected system  $\Sigma\!=\!$ $\mathcal{I}(\Sigma_1,\Sigma_2)$ depicted in Figure~\ref{fig:smallgain}, consisting of two subsystems $\Sigma_1$ and $\Sigma_2$ each described in \eqref{sys:linear}.
For each $\Sigma_i$, we construct a local finite abstraction $\hat \Sigma_i=( \hat X_i, \hat X_{0_i}, \hat X_{S_i},  \varnothing, \hat W_i,\hat f_i,\hat X_i, \mathrm{id})$  as in \eqref{symbolicmodel} via so-called local approximate initial-state opacity-preserving simulation functions $V_i:X_i\times\hat X_i\to\R_{\geq0}$ satisfying the following conditions: 
\begin{enumerate}[leftmargin=*]
	\item\begin{enumerate}
	\item $\forall x_{0} \! \in\!  { X}_{0_i} \! \cap\!  { X}_{S_i}$, $\exists \hat x_{0_i} \in \hat { X}_{0_i} \! \cap\!  \hat { X}_{S_i}$, s.t. $V_{i}(x_{0_i},\!\hat x_{0_i})  \leq  \epsilon_i$; 
	\item $\forall \hat x_0  \in  \hat { X}_{0_i}   \! \setminus\!   \hat { X}_{S_i}$, $\exists x_{0_i}  \in   { X}_{0_i}   \! \setminus \! { X}_{S_i}$, s.t. $V_{i}(x_{0_i},\hat x_{0_i})   \leq   \epsilon_i $;
		\end{enumerate}
	\item[2] $\forall x_i \in  X_i, \forall \hat x_i \in \hat { X}_i$, $\Vert x_i - \hat x_i \Vert \leq V_{i}(x_i,\hat x_i)$;
	\item[3] $\forall x_i \!\in\!  X_i, \forall \hat x_i \!\in\! \hat { X}_i$ s.t. $V_{i}(x_i,\hat x_i) \!\leq\! \epsilon_i$, $\forall w_i \!\in\!  W_i$, $\forall \hat w_i \!\in\! \hat{{W}}_i$ s.t. $\Vert w_i\!-\! \hat w_i \Vert \!\leq\! \vartheta_i$, the following  hold: 
		\begin{enumerate}
			\item $\forall  {x}_i^+ $,  $\exists  \hat{x}_i^+$, s.t. $V_{i}({x}_i^+, \hat{x}_i^+) \leq \epsilon_i$;
			\item  $\forall  \hat{x}_i^+$,  $\exists {x}_i^+$,  s.t. $V_{i}({x}_i^+,\hat{x}_i^+) \leq \epsilon_i$,
		\end{enumerate}
	\end{enumerate}
where $\epsilon_i, \vartheta_i \in \mathbb{R}_{\geq 0}$.
If $\gamma_{1}\gamma_{2} <1$ (similar to the small gain criterion in \cite{zames1966input}), then $V(x,\xhat) = \max_{i = 1,2} \{V_i(x_i, \xhat_i) \}$ is an approximate initial-state opacity-preserving simulation function from $\mathcal{I}(\Sigma_1,\Sigma_2)$ to $\mathcal{I}(\hat \Sigma_1,\hat \Sigma_2)$.
\end{theorem}
\end{tcolorbox}


\begin{figure}[t!]
    \centering
 	\includegraphics[scale=0.8]{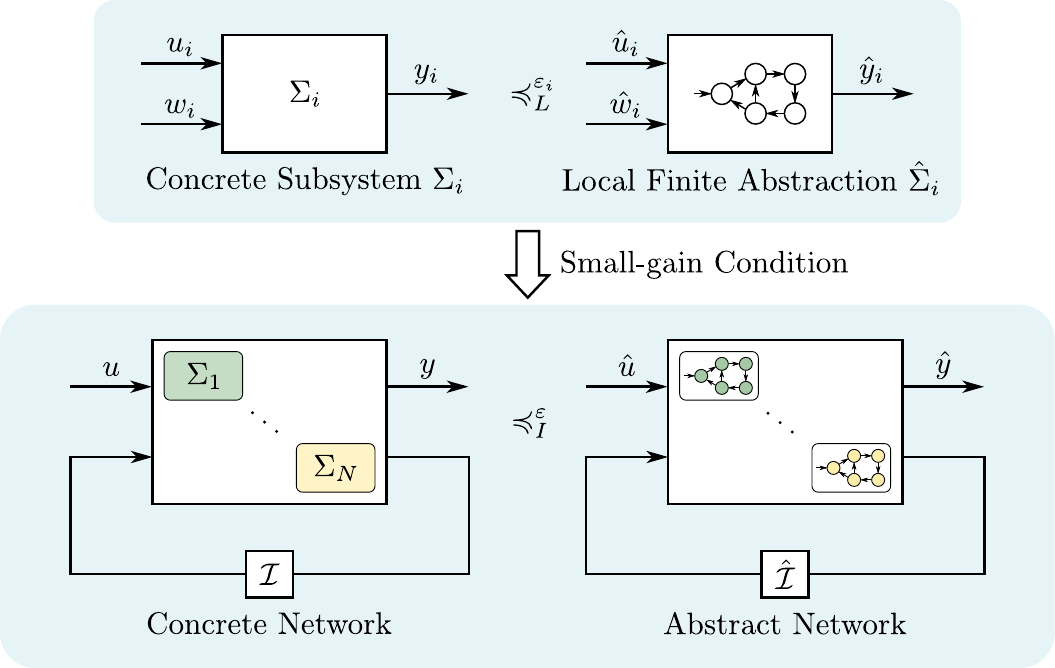}
    \caption{Compositional framework for the construction of opacity-preserving finite abstractions for interconnected systems.}
    \label{fig:compo_scheme}
    \vspace{-0.3cm}
\end{figure}

Note that similar results can be obtained for interconnections of $N$ subsystems with general dynamics as shown in \cite{liu2021compositional}. 
More details can be found there on the compositionality results tailored to different types of opacity as well. 

As can be observed from the theorem, the compositional framework is based on a  small-gain type condition.
Small-gain theorems have a long-known history in control design dating back to the 1960's \cite{zames1966input}. They have been extensively leveraged to establish stability properties of interconnected systems  \cite{jiang1994small,dashkovskiy2007iss}.
In our work, the small-gain type condition is imposed on the concrete network of subsystems for the existence of proper compositional finite abstractions. More specifically, it facilitates the compositional construction of finite abstractions by certifying a small (weak) interaction of the subsystems which prevents an amplification of the signals across the possible interconnections.

The intuition behind the proposed compositionality result is as follows.
Instead of tackling the overall system in a monolithic manner, the compositional scheme provided here allows us to build an abstraction for the overall system by dealing with subsystems only. In particular, new notions of approximate opacity-preserving simulation functions are first introduced for both subsystems and the interconnected system, which provide the basis for using abstraction-based techniques in verifying approximate opacity for large-scale interconnected systems. Based on the local simulation functions, one can construct local finite abstractions for subsystems individually. Then, under a small-gain type condition, a compositionality result is derived which ensures that the interconnection of local abstractions mimics the behavior of the concrete interconnected system in terms of preserving opacity.
An algorithm (\cite[Algorithm 1]{liu2021compositional}) is provided as a guideline to design quantization parameters of local finite abstractions. 
The compositionality scheme proposed in this paper is schematically illustrated in Figure.~\ref{fig:compo_scheme}.

\subsection{Modular Verification for Large-scale CPS: A Barrier Certificate Approach}

As presented in Section.~\ref{sec:veri_bc}, barrier certificates can be leveraged as an useful alternative approach for the verification of opacity for CPS. 
Though promising, the computation of such types of barrier certificates is still an expensive problem, which may become intractable while dealing with large-scale interconnected systems. 
In this subsection, we briefly describe the recent results developed in \cite{kalat2021modular} for a compositional approach for verifying approximate opacity via the construction of barrier certificates. This result shows that by employing a small-gain type condition, a barrier certificate  for an interconnected system  as in Theorem \ref{BC} can be constructed by composing so-called local barrier certificates of subsystems. 

Let us again consider the feedback interconnection of two subsystems $\Sigma_1$ and $\Sigma_2$ each described as in \eqref{sys:linear} and associated with gain functions $\gamma_{i} = \vert b_i/(1-a_i)\vert$. 
The main compositionality result proposed in \cite{kalat2021modular} is summarized as follows.

\begin{tcolorbox}
\begin{theorem}[\textbf{Compositional Construction of Barrier Certificates for Verifying Opacity}] \label{Main:compobc}\upshape
Consider the interconnected system  $\Sigma\!=\!$ $\mathcal{I}(\Sigma_1,\Sigma_2)$  depicted in Figure~\ref{fig:smallgain}, consisting of two subsystems $\Sigma_1$ and $\Sigma_2$ each described in \eqref{sys:linear}. For each $\Sigma_i$, we construct a so-called local barrier certificate $B_i:X_i\times X_i\to\R$ for the augmented subsystem $\Sigma_i \times \Sigma_i$ satisfying the following conditions
\begin{align*}
  &\forall (x_i,\xhat_i) \in \mathcal{R}_i,  \quad \quad \quad \quad \quad B_i(x_i,\xhat_i) \geq  \Vert(x_i,\xhat_i)\Vert, \\ 
     &\forall (x_i,\xhat_i) \in \mathcal{R}_{0i}, \quad \quad \quad \quad \quad B_i(x_i,\xhat_i) \leq 0, \\ 
      &\forall (x_i,\xhat_i) \in \mathcal{R}_{ui}, \quad \quad \quad \quad \quad B_i(x_i,\xhat_i) > 0, \\
    &\forall (x_i,\xhat_i) \in \mathcal{R}_i,~~\forall (x_j,\xhat_j) \in \mathcal{R}_j, \\
    & \quad \quad \quad \quad B_i({x}_i^+,  \hat{x}_i^+) \leq   (1-a_i)B_i(x_i,\xhat_i)+b_{i}\Vert(x_j,\xhat_j)\Vert,
\end{align*}
where sets $\mathcal{R}_{0i}$ and $\mathcal{R}_{ui}$ are the projections of sets $\mathcal{R}_{0}$ and $\mathcal{R}_{u}$ as in \eqref{set:initial}-\eqref{set:unsafe} over the augmented subsystem $\Sigma_i \times \Sigma_i$. 
If  $\gamma_{1}\gamma_{2} <1$  holds,
then  $B(x,\xhat) = \max_{i = 1,2} \{B_i(x_i, \xhat_i) \}$ is a barrier certificate for the augmented interconnected system $\Sigma \times \Sigma$, which implies that the interconnected system $\Sigma$ is $\delta$-approximate initial-state opaque.
\end{theorem}
\end{tcolorbox}


Note that local barrier certificates of subsystems are mainly used for constructing overall barrier certificates for the interconnected systems, and they are not useful on their own to verify opacity properties.
The above results show that, under a small-gain type condition, a barrier certificate $B$ for the augmented interconnected system can be obtained by composing local barrier certificates computed for subsystems.
As presented in Sec.~\ref{sec:veri_bc}, if we can find a barrier certificate for the interconnection of augmented subsystems, one obtains that the original large-scale interconnected system is approximately initial-state opaque.  
Note that similar results can be obtained for interconnections of $N$ subsystems with general dynamics as shown in \cite{kalat2021modular}. The compositional construction of barrier certificates which implies the lack of opacity (as in Theorem.\ref{BC1}) of large CPS can be achieved by a similar framework as well.



\subsection{Ongoing \& Open Problems}
Here, we mention some potential future directions on compositional approaches for opacity verification and synthesis.

\paragraph{\emph{Efficient Models for Concurrent Systems}}
Interconnected systems are inherently concurrent, for which the  major computational challenge  comes from the issue of state-space explosion. For discrete systems, instead of using labeled transition systems, many alternative models have been proposed to efficiently represent large-scale concurrent
systems without enumerating the composed state space; one of the most widely used models is Petri nets \cite{cassandras2021introduction}.   
Using Petri nets as the underlying model for opacity verification goes back to the seminal work of Bryans et al.~\cite{bryans2008opacity}. 
Unfortunately, it has been proved that opacity verification is generally undecidable for unbounded Petri nets \cite{tong2017decidability,berard2018complexity,masopust2019deciding}. 
On the other hand, for bounded Petri nets, many computationally efficient approaches have been developed recently by utilizing   structural properties and modularity of Petri nets to overcome the issue of state-space explosion; see, e.g., \cite{ma2017basis,tong2017verification,cong2018line,saadaoui2020current,lefebvre2020privacy, tommasi2021optimization}. However, all these results can only be applied to finite systems. How  to abstract concurrent interconnected CPS using Petri nets while preserving opacity properties is an interesting future direction. 

\paragraph{\emph{Leverage Existing Modular Algorithms}}
In the past decades, despite those opacity-related modular techniques already mentioned in Section~\ref{subsec:modular},  there are already numerous different modular verification and synthesis methods  developed for other non-security properties in DES and formal methods literature. For example, in the context of supervisory control of DES, researchers have proposed many effective modular controller synthesis approaches using, for example,   
state tree structures \cite{ma2006nonblocking,chao2013modular}, 
hierarchical interfaces
\cite{leduc2005hierarchical,hill2010multi}, 
multi-level coordinators 
\cite{komenda2015coordination}, and equivalence-based abstractions \cite{feng2008supervisory,su2010model}. 
There are also numerous recent works exploring the philosophy  of compositional reasoning in the context of reactive synthesis; see, e.g.,  
\cite{alur2018compositional,majumdar2020assume,bakirtzis2021compositional}.
We believe that  many of the aforementioned modular/compositional approaches for non-security properties can be generalized to incorporate the security constraints, which deserve deeper and detailed investigations.

\paragraph{\emph{Distributed Secure-by-Construction Synthesis}}
For large-scale interconnected systems, the abstract interconnection constitutes several relatively smaller
local finite abstractions, as investigated in  Section~\ref{sec:compo_abs}, that run synchronously. Since the controller synthesis problem for LTL specifications has severe worst-time complexity (doubly exponential), computing the monolithic product of
all of the finite components makes the synthesis highly impractical. Moreover, often it may be impractical to
assume that subsystems have complete knowledge of the states of other subsystems. To model
these scenarios, one can represent the system as a network of finite abstractions where each subsystem has
a separate mission and opacity requirement. Some of the states of neighbouring local finite abstractions may be
shared with other local abstractions. This gives rise to the distributed reactive synthesis problem \cite{Schewe08} where the system
consists of several independent processes that cooperate based on local information to accomplish a global
specification. Such a setting changes the synthesis problem from a two-player complete-information game
to two-player games of incomplete information \cite{Reif84}. However, even for safety and reachability objective
(sub-classes of LTL), it is well known \cite{PR90,schewe14}  that the distributed synthesis problem is undecidable for
general interconnected systems. There are two directions to achieve decidability: the first is to restrict the
network architecture \cite{PR90} and the second is the approach of bounded synthesis \cite{Schewe07} as we have already discussed for the case of monolithic sysnthesis. 
\section{Future Directions}
Next, we touch upon some potential directions related to the overall secure-by-construction theme that differ from the parameters of study in this technical introduction. 
We believe that these directions may provide impetus to  research in security-critical system design.
 


\subsection{Information-Theoretic Foundations} 
The concept of privacy discussed so-far in this paper is binary: either a system leaks information or it does not leak any information. 
However, in practice such binary mitigation may not be feasible and may require an information-theoretic prospective on quantifying and minimizing the amount of information leak. 
Shannon, in his seminal paper~\cite{shannon1948mathematical}, coined and popularized the notion of \emph{entropy} in measuring information: for a random variable $X$ with values in some domain $\mathcal{X}$, the entropy of (or the uncertainty about) $X$, denoted by $H(X)$, is defined as 
\[
H(X) = \sum_{x \in \mathcal{X}} P[X= x] \log_2 \frac{1}{P[X=x]}.
\]
Shannon proved that $H(X)$ is the only function (modulo scaling) that satisfies the natural continuity, monotonicity, and choice decomposition (See~\cite{shannon1948mathematical}, for more details).
Similarly, for jointly distributed random variables $X$ and $Y$, the conditional entropy $H(X\mid Y)$, i.e. uncertainty about $X$ given $Y$, can be defined as 
\[
H(X \mid Y) = \sum_{y \in \mathcal{Y}} P[Y= y] H(X \mid Y = y),
\]
where $\mathcal{Y}$ is the domain of $Y$. These definitions provide us a way to measure the \emph{information loss}: if $H(X)$ is the uncertainty about $X$ and if $H(X\mid Y)$ is the uncertainty about $X$ after $Y$ is revealed, the information loss in this process is $I(X; Y) = H(X) - H(X \mid Y)$. 
Smith~\cite{smith2009foundations} introduced an alternative notion of entropy called the \emph{guessing entropy} $G(X)$ that corresponds to the number of guesses required to infer the value of $X$: of course a rational strategy in guessing these values will be to guess them in a non-increasing sequence of probability, hence $G(X) = \sum_{i=1}^{n} i p_i$ where $\seq{p_1, p_2 ,\ldots, p_n}$ is the sequence of probabilities of elements of $X$ arranged in an non-increasing fashion.

The notion of opacity discussed in this paper requires that the attacker should deduce nothing about all opacity properties of the system from observing the outputs of the system.  However, achieving full opacity may not be possible in general, because oftentimes systems reveal information depending on the secret properties.
To extend the notion of opacity to quantitative opacity, we can use the quantitative notion of information leakage.
We say that two opacity properties $\alpha, \alpha'$ are {\it indistinguishable} in $\Sigma$, and we write $\alpha \equiv_\Sigma \alpha'$,  if for any trace $r$ satisfying $\alpha$, there exists another trace $r'$ satisfying $\alpha'$ such that both $r$ and $r'$ have analogous observations, i.e. $h(r)=h(r')$. 
Let us generalize the original set of opacity properties from $\set{\alpha, \neg \alpha}$ to $\overline{\alpha}= \set{\alpha_1, \ldots, \alpha_n}$. In this case, the system $\Sigma$ is called {\it opaque}, if every pair of opacity properties in $\overline{\alpha}$ are mutually indistinguishable.
Let $Q= \{Q_1, Q_2, \ldots, Q_k\}$  be the quotient space of $O$
characterized by the indistinguishability relation. 
Let $B_Q = \seq{B_1, B_2, \ldots, B_k}$ be the sizes of observational
equivalence classes from $Q$; let $B = \sum_{i=1}^k B_i$.
Assuming uniform distributions on $Q$, K\"opf and Basin~\cite{KB07} characterize expressions for various information-theoretic measures on information leaks which are given below:
\begin{enumerate}[leftmargin=*]
\item {\it Shannon Entropy:}
  $SE(\Sigma,\overline{\alpha}) = (\frac{1}{B})  \sum\limits_{1 \leq i \leq k} B_i \log_2(B_i)$,
\item {\it Guessing Entropy:}
  $GE(\Sigma,\overline{\alpha}) = (\frac{1}{2B}) \sum\limits_{1 \leq i \leq k} B_i^2 + \frac{1}{2}$,
 \item {\it Min-Guess Entropy:}
  $MG(\Sigma,\overline{\alpha}) {=} \min\limits_{1 \leq i \leq k} \set{(B_i + 1)/2}$.
\end{enumerate}

This allows us to generalize our opacity requirements in a quantitative fashion. 
Given a property  $\varphi$ as a mission requirement, and
 opacity property tuple $\overline{\alpha} = \{\alpha_1, \ldots, \alpha_k\}$, an entropy bound $K$ and the corresponding entropy criterion $\kappa \in \set{SE, GE, MG}$, the quantitative security-aware
 verification $\Sigma \models (\varphi, \overline{\alpha})$ is to decide whether $\Sigma \models \varphi$ and $\kappa(\Sigma,\overline{\alpha}) \leq K$.
Similarly, the quantitative security-aware synthesis is to design a supervisor/controller $C$ such that  
$\Sigma_C \models (\varphi, {\overline{\alpha}})$.

Quantitative theory of information have been widely used
for the verification of security properties~\cite{smith2009foundations,KB07,backes2009automatic,heusser2010quantifying} in the context of finite state and software systems.
Moreover, for such systems several 
restricted classes of synthesis approaches~\cite{kopf2009provably,askarov2010predictive,zhang2011predictive,zhang2012language,kadloor2012mitigating,schinzel2011efficient,DBLP:conf/cav/Tizpaz-NiariC019} have been proposed that focus on side-channel  mitigation techniques by increasing the remaining entropy of secret sets leaked while maintaining the performance.

\subsection{Data-Driven Approaches for CPS Security}
This paper assumed the access to a model of the system and proposed security-aware verification and synthesis approaches.
Oftentimes, a true explicit model of the system is not available or is too large to reason with formally. 
Reinforcement learning~\cite{sutton2018reinforcement} (RL) is a sampling-based optimization algorithm that computes optimal policies driven by scalar reward signals. 
Recently, RL has been extended to work with formal logic~\cite{LTLBased2,camacho2019ltl,oura2020reinforcement,hasanbeig2019certified,lavaei2020formal}, and automatic structures ($\omega$-automata~\cite{hahn2019omega,mateo21} and reward machines~\cite{RM1}) instead of scalar reward signals. 
A promising future direction is to extend RL-based synthesis to reason with security properties of the system. 

The controller learned via deep RL will have deep neural networks as the controllers. 
Additionally, deep neural networks are often employed in place of cumbersome tabular controllers to minimize the size of the program logic. 
In such systems, security verification need to reason with neural networks along with the system dynamics. 
There is a large body of work~\cite{huang2017safety,Abate21,hahn2019omega,pulina2012challenging,lomuscio2017approach,xiang2018reachability,lavaei2020formal} in verifying control systems with neural networks using SMT solvers, and will provide a promising avenue of research in developing security verification and synthesis approaches for CPS with neural networks based controllers.


Radical advances in inexpensive sensors, wireless technology, and the Internet of Things (IoT) offer unprecedented opportunities by ubiquitously collecting data at high detail and at large scale. Utilization of data at these scales, however, poses a major challenge for verifying or designing CPS, particularly in view of the additional inherent uncertainty that data-driven signals introduce to systems behavior and their correctness. In
fact, this effect has not been rigorously understood to this date, primarily due to the missing link between
data analytics techniques in machine learning/optimization and the underlying physics of CPS. A future research direction is to
develop scalable data-driven approaches for formal verification and synthesis of CPS with unknown closed form
models (a.k.a. black-box systems) with respect to both mission and security properties. The main novelty is to bypass the model identification phase
and directly verify or synthesize controller for CPS using system behaviors. The main reasons behind the quest to directly work on system behaviors and bypass the identification
phase are: i) Identification can introduce approximation errors and have a large computational complexity; ii) Even when
the model is known, formal verification and synthesis of CPS are computationally challenging.

\subsection{Security for Network Multi-Agent CPS}
This paper mostly discussed a centralized setting for CPS security, i.e., a single CPS plant with global secrets against a single attacker, although the CPS itself  may consist of several smaller subsystems.   
However, in many modern engineering systems such as connected autonomous vehicles \cite{lu2014connected}, smart micro-grids \cite{yu2016smart} and smart cities \cite{cassandras2016smart}, there may exist no centralized decision-maker. 
Instead, each CPS agent  interacts and collaborates/competes with each other via information exchanges over networks to make decisions, which leads to the network multi-agent CPS.  There is a large body of  works~\cite{guo2015multi,tumova2016multi,guo2016communication,kantaros2016distributed,schillinger2018simultaneous,sahin2019multi} in synthesizing coordination strategies for network multi-agent CPS  for high-level mission requirements using formal methods. However, the security issue, which is more severe in multi-agent CPS due to large communications and information exchanges, is rarely considered.   
In particular, in multi-agent CPS, each agent may have its own security considerations that depend on the time-varying configurations of the entire network. Therefore, how to  define formal security notions  that are suitable for multi-agent systems is an important but challenging future direction. 

Recently, security and privacy considerations over networks have attracted significant attentions in the context of distributed state estimations \cite{mitra2019byzantine,an2022enhancement}, distributed averaging/consensus \cite{mo2017privacy,hadjicostis2020privacy}, distributed optimizations \cite{han2017differentially,lu2018privacy}, and distributed machine learning \cite{huang2020instahide,li2020federated}.  
However, those results are mostly developed for distributed computing systems and are not directly applicable for multi-agent CPS with heterogeneous dynamics. Furthermore, most of the existing security-aware protocols for distributed systems are designed for specific tasks and there is still a lack of formal methodologies for security-aware verification and secure-by-construction synthesis of  communication protocols and coordination strategies for network multi-agent CPS. 
Finally, rather than a single passive attacker, network CPS may suffer from  multiple  active malicious attackers.   
Therefore, one needs to develop effective approaches for characterizing and controlling the evolution of security properties over dynamic networks of multiple players. A promising future direction is to develop a comprehensive framework for multi-agent CPS security by extending formal reasoning with multi-player game-theory. 

\section{Conclusion}
This paper may serve as an excursion into some prominent ideas and formalism from three distinct fields of formal methods, discrete-event systems, and control theory to study secure-by-construction synthesis paradigm. 
We intentionally kept the technical discussion at a higher-level to expand the readership and aimed to provide necessary background and references, where appropriate. 
We synthesized a general setting of security-aware verification and secure-by-construction synthesis integrating various notions of privacy and correctness in a common framework. 
While this article is primarily informed by the research interests of the authors, we hope that it provides the basic foundations on which the related questions can be posed and answered. 

We shall draw the readers' and potential researchers' attention that, security has been a moving goalpost and more
damaging vulnerabilities are yet unknown. The proposed
approaches in this paper need to be combined with classical fuzzing-based security
research to uncover previously undiscovered security vulnerabilities.
Moreover, most of the existing results on security analysis for CPS remain mainly theoretical. 
Over the past few years, several software tools (e.g., \texttt{DESUMA} \cite{ricker2006desuma}, \texttt{SUPREMICA} \cite{akesson2006supremica}, and \texttt{TCT} \cite{feng2006tct}) have been developed for the analysis of DES modeled as finite automata, which are shown to be useful in the verification or synthesis of opacity properties for finite systems.
Our prior research has produced software tools including  \texttt{SCOTS} \cite{scots}, \texttt{pFaces} \cite{pfaces}, \texttt{OmegaThreads} \cite{omegathread}, \texttt{DPDebugger} \cite{tizpaz2018differential} and \texttt{Schmit} \cite{DBLP:conf/cav/Tizpaz-NiariC019}, which provides formal, automated
abstractions of complex CPS and of reactive synthesis.
There is a great need to develop efficient toolboxs and proof-of-concept benchmarks to evaluate the practical feasibility of the foundations and algorithms developed for
abstracting, analyzing, or enforcing security properties over complex CPS. 
In addition to academic benchmarks, it is important to improve the applicability of theoretical methods to industrial case studies and real-life applications. Designing open access courses that provide an ``end-to-end
view", starting from the foundations of control and discrete systems theory and going into security issues for CPS is also needed to train students, particularly those deciding to pursue research or work
professionally on autonomous systems.


\bibliographystyle{abbrv}
\bibliography{mybibfile}









\end{document}